\documentclass[review]{elsarticle}
\usepackage[linesnumbered,ruled,vlined]{algorithm2e}
\usepackage{graphicx}
\usepackage{amssymb}
\usepackage{amsmath}
\usepackage{amsfonts}
\usepackage{mathtools}
\DeclarePairedDelimiter\ceil{\lceil}{\rceil}

\newtheorem{mythm}{Theorem}
\usepackage{hyperref}
\hypersetup{
colorlinks=true,
linkcolor=blue,
filecolor=magenta,      
urlcolor=cyan,
}
\usepackage{xcolor}

\usepackage[framemethod=TikZ]{mdframed}
\mdfdefinestyle{MyFrame}{%
	linecolor=blue,
	outerlinewidth=1.0pt,
	roundcorner=20pt,
	innertopmargin=\baselineskip,
	innerbottommargin=\baselineskip,
	innerrightmargin=20pt,
	innerleftmargin=20pt,
	backgroundcolor=gray!50!white}
\usepackage{multirow}

\journal{Elsevier}

\begin{document}

\begin{frontmatter}

\title{A Survey on Influence Maximization in a \\ Social Network}

\author[mymainaddress]{Suman Banerjee}
\ead{suman@iitkgp.ac.in}
\author[mymainaddress]{Mamata Jenamani}
\ead{mj@iem.iitkgp.ac.in}
\author[mymainaddress]{Dilip Kumar Pratihar}
\cortext[mycorrespondingauthor]{Corresponding author-Dilip Kumar Pratihar}
\ead{ dkpra@mech.iitkgp.ac.in}

\address[mymainaddress]{Indian Institute of Technology, Kharagpur, West Bengal, India.}

\begin{abstract}
Given a \textit{social network} with \textit{diffusion probabilities} as edge weights and an integer $k$, which $k$ nodes should be chosen for initial injection of information to maximize influence in the network? This problem is known as \textit{Target Set Selection in a social network} (\textit{TSS Problem}) and more popularly, \textit{Social Influence Maximization Problem} (\textit{SIM Problem}). This is an active area of research in \textit{computational social network analysis} domain since one and half decades or so. Due to its  practical importance in various domains, such as \textit{viral marketing}, \textit{target advertisement}, \textit{personalized recommendation}, the problem has been studied in different variants, and different solution methodologies have been proposed over the years. Hence, there is a need for an organized and comprehensive review on this topic. This paper presents a survey on the progress in and around  \textit{TSS Problem}. At last, it discusses current research trends and future research directions as well.
\end{abstract}

\begin{keyword}
Target Set Selection Problem \sep Social Networks \sep Influence Maximization \sep Inapproxibility Results \sep  Approximation Algorithm \sep Greedy Strategy \sep NP\mbox{-}Hard Problem.
\MSC[2010] 00-01\sep  99-00
\end{keyword}

\end{frontmatter}


\section{Introduction}

A social network is an interconnected structure of a group of agents formed for social interactions \cite{liu2011social}. Nowadays, social networks play an important role in spreading information, opinion, ideas, innovation, rumors etc. \cite{centola2010spread} \cite{nekovee2007theory}. This spreading process has a huge practical  importance in viral marketing \cite{leskovec2007dynamics} \cite{chen2010scalable}, personalized recommendation \cite{song2006personalized}, feed ranking \cite{ienco2010meme}, target advertisement \cite{li2015real}, selecting influential twitters \cite{weng2010twitterrank} \cite{bakshy2011everyone}, selecting informative blogs \cite{leskovec2007cost}, etc. Hence, recent years have witnessed a significant attention in the study of \textit{influence propagation} in online social networks. Consider the case of viral marketing of a commercial house, where the goal is to attract the users for purchasing a particular product. The best way to do this is to select a set of highly influential users and distribute them free samples. If they like the product, they will share the information to their neighbors. Due to their high influence, many of the neighbors will try for the product and share the information to their neighbors. This cascading process will be continued and ultimately a large fraction of the users will try for the product. Naturally, number of free sample products will be limited due to economic reason. Hence, this process will be fruitful, if the free samples can be distributed among the highly influential users and the problem here bottoms down to select influential users from the network. This problem is known as \textit{Social Influence Maximization Problem} \footnote{Now onwards, we will use Target Set Selection and Social Influence Maximization interchangeably}. 
\par Social influence occurs due to the diffusion of information in the network. This phenomenon in a networked system is well studied \cite{cowan2004network} \cite{kasprzak2012diffusion}. Specifically, there are two popularly adopted models to study the diffusion process, namely \textit{Independent Cascade Model} (abbreviated as \textit{IC Model}), which collects the independent behavior of the agents, and the other one is \textit{Linear Threshold Model} (abbreviated as \textit{LT Model}), which captures the collective behavior of the agents (detailed discussion is deferred till  Section \ref{BID}) \cite{shakarian2015independent}. In both the models, information is diffused in discrete time steps from some initially identified nodes and continued for several rounds. In SIM Problem, our goal is to maximize influence by selecting appropriate seed nodes.
\par To study the SIM Problem, a social network is abstracted as a \textit{graph} with the users as the \textit{vertex set} and \textit{social ties} among the users as the edge set. It is also assumed that the \textit{diffusion threshold} (a measurement of how hard to influence the user and given in a numerical scale; more the value, more hard to influence the user) is given as the \textit{vertex weight} and influence probability between two users as \textit{edge weight}. In this settings, the SIM Problem is stated as follows: for a given size $k$ ($k \in \mathbb{Z}^{+}$), choose the set $\mathcal{S}$ of $k$ nodes, such that $\sigma(\mathcal{S})$ gets maximized \cite{sun2011survey}. Here $\sigma(.)$ is the \textit{social influence function}. For any given seed $\mathcal{S}$, $\sigma(\mathcal{S})$ returns the set of influenced nodes, when the diffusion process is over.

\subsection{Focus and Goal of the Survey} 
\begin{figure}
\centering
  \includegraphics[scale=0.25]{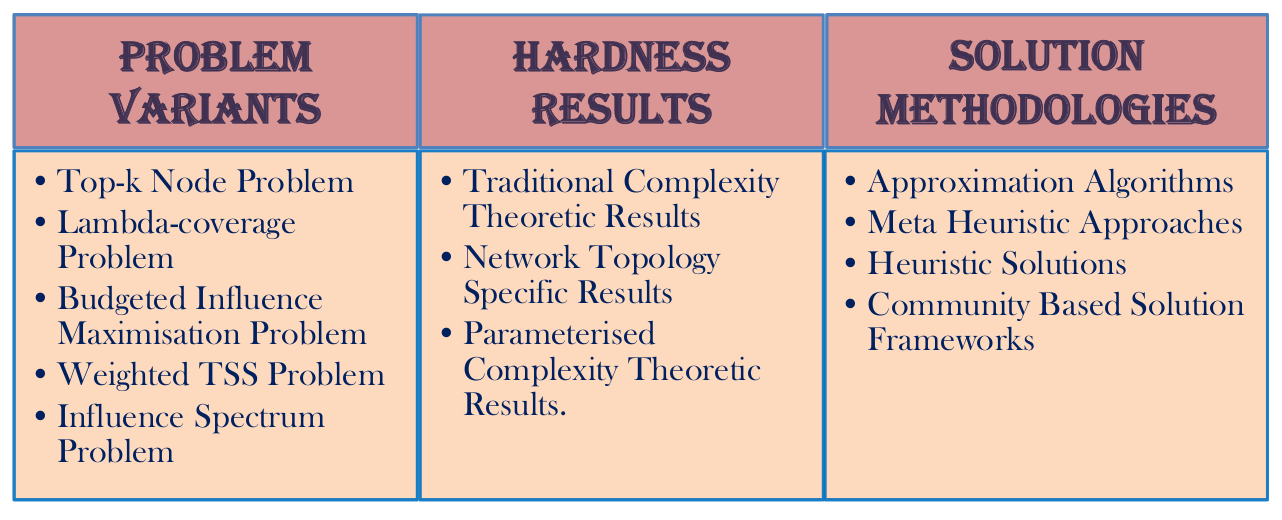}
  \caption{Overview of this survey}
  \label{fig:1}
\end{figure}

In this survey, we have mainly focused on three aspects of the problem, as mentioned below.
\begin{itemize}
\item Variants of this problem studied in the literature,
\item Hardness results of this problem in both traditional as well as parameterized complexity framework,
\item Different solution approaches proposed in the literature.
\end{itemize}
The overview of this survey is shown in Figure \ref{fig:1}. There are several other aspects of the problem, such as \textit{SIM in the presence of adversaries}, \textit{in a time\mbox{-}varying social network}, \textit{in competitive scenario} etc., which we have not considered in this survey.

 The main goal of this survey is threefold:
 \begin{itemize}
 \item to provide comprehensive understanding about the SIM Problem and its different variants studied in the literature,
 \item to develop a taxonomy for classifying the existing solution methodologies and present them in a concise manner,
 \item to present an overview of the current research trend and future research directions regarding this problem. 
\end{itemize}
We set the following two criteria for the studies to be included in this survey:
\begin{itemize}
\item Research work presented in the publication should produce theoretically or empirically better than some of the previously published results. 
\item The presented solution methodology should be generic, i.e., it should work for a network of any topology.    
\end{itemize}   
\subsection{Organization of the Survey}
\par Rest of the paper is organized as follows: Section \ref{Sec:Bac} describes some background material required to understand the subsequent sections of this paper. Section \ref{Sec:VTSSP} formally introduces the SIM Problem and its variants studied in the literature. Section \ref{Sec:Hard} describes hardness results of this problem in both traditional as well as parameterized complexity theory framework. Section \ref{Sec:MRC} describes some major research challenges in and around this problem. Section \ref{Sec:SolTSS} describes the proposed taxonomy for classifying the existing solution methodologies in different categories and discuss them. Section \ref{Sec:SRD} presents the summary of the survey and gives some future research directions. Finally, Section \ref{CR} presents concluding remarks regarding this survey.

\section{Background} \label{Sec:Bac}
In this section, we have described relevant background topics upto required depth, such as \textit{basic graph theory}, relation between SIM and existing graph theoretic problems, \textit{approximation algorithm}, \textit{parameterized complexity theory} and \textit{information diffusion models} in social networks. The symbols and notations that have been used in the subsequent sections of this paper are given in Table \ref{Tab : 1}.
\begin{table} \label{Tab:1}
 	\centering
 	\caption{Symbols and Notations}
 	\label{Tab : 1}
 	\begin{tabular}{|c|c|}
 		\hline
 		{\ \textbf{Symbols}}  & {\ \textbf{Interpretation}}\\
 		\hline
 		$G(V, E, \theta, \mathcal{P})$ &  Directed, vertex and edge weighted social network\\
 		$V(G)$ & Set of vertices of network $G$\\
 		$E(G)$ & Set of edges of network $G$\\
 		$U$ &  Set of users of the network, i.e., $U=V(G)$\\
 		$n$ & Number of users of the network, i.e., $n=\vert V(G) \vert$\\
 		$m$ & Number of Edges of the network, i.e., $m=\vert E(G) \vert$\\
 		$\theta$ &  Vertex weight function of $G$, i.e., $\theta:V(G) \longrightarrow [0,1]$\\
 		$\theta_i$    & Weight of vertex $u_i$, i.e., $\theta_i=\theta(u_i)$\\
 		$\mathcal{P}$	     &  Edge weight function, i.e., $\mathcal{P}:E(G) \longrightarrow [0,1]$\\
 		$p_{ij}$  & Edge weight of the edge $(u_iu_j)$\\
 		$\mathcal{N}(u_i)$ & Open neighborhood of vertex $u_i$\\
 		$\mathcal{N}[u_i]$ & Closed neighborhood of vertex $u_i$\\
 		$[n]$      & Set $\left\{1, 2,\dots\ , n \right\}$\\
 		$\mathcal{N}^{in}(u_i)$  & Incomming neighbors of vertex $u_i$ \\
 		$\mathcal{N}^{out}(u_i)$ & Outgoing neighbors of vertex $u_i$\\
 		$deg^{in}(u_i)$ & Indegree of vertex $u_i$\\
 		$deg^{out}(u_i)$ & Outdegree of vertex $u_i$\\
 		$dist(u,v)$ & Number of edges in the shortest path between $u$ and $v$.\\
 		$\mathcal{S}$ & Seed set for diffusion, i.e., $\mathcal{S} \subset V(G)$\\
 		$k$ & Maximum allowable cardinality for the seed set, i.e.,  $\vert \mathcal{S} \vert \leq k$\\
 		$r$ & Maximum allowable round for diffusion \\
 		\hline
 	\end{tabular}
 \end{table}
\subsection{Basic Graph Theory} \label{BGT}
 Graphs are popularly used to represent most of the real world networked systems including social networks \cite{campbell2013social} \cite{wang2011understanding}. Here, we have reported some preliminary concepts of \textit{basic graph theory} from \cite{diestel2005graph}. A graph is denoted by $G(V, E)$ where $V(G)$ and $E(G)$ are the \textit{vertex set} and \textit{edge set} of $G$, respectively. For any arbitrary vertex, $u_i \in V(G)$, its \textit{open neighborhood} is defined as $\mathcal{N}(u_i)=\left\{u_j \vert (u_iu_j) \in E(G)\right\}$. \textit{Closed neighborhood} of $u_i$ will be $\mathcal{N}[u_i]=u_i \cup \mathcal{N}(u_i)$. \textit{Degree} of a vertex is defined as the \textit{cardinality} of its open neighborhood, i.e., $deg(u_i)=\vert \mathcal{N}(u_i) \vert$. For any $S \subset V(G)$, its open neighborhood and close neighborhood will be $\mathcal{N}(S)=\underset{u_i \in S}{\cup} \mathcal{N}(u_i)$ and $\mathcal{N}[S]=S \cup \mathcal{N}(S)$, respectively. Two vertices $u_i$ and $u_j$ are said to be \textit{true twins}, if $\mathcal{N}[u_i]=\mathcal{N}[u_j]$ and \textit{false twins}, if $\mathcal{N}(u_i)=\mathcal{N}(u_j)$. A graph is  \textit{weighted}, if a real number is associated with its vertices or edges or both. A graph is \textit{directed}, if its edges have directions. The edges that join the same pair of vertices are known as parallel edges, and an edge whose both the end points are same is known as \textit{self\mbox{-}loop}. A graph is \textit{simple}, if it is free from self\mbox{-}loop and parallel edges. 

\par Information diffusion process in a \textit{social network} is represented by a \textit{simple}, \textit{directed} and \textit{vertex} and \textit{edge weighted graph} $G(V, E, \theta, \mathcal{P})$. Here, $V(G)= \left\{u_1, u_2,\dots\ , u_n\right\}$, the set of users of the network and $E(G)= \left\{e_1, e_2,\dots\ , e_m\right\}$, the set of social ties among the users. $\theta$ and 
$\mathcal{P}$ are the \textit{vertex} and \textit{edge weight} function, which assign a numerical value in between $0$ and $1$ to each vertex and edge, respectively, as its weight, i.e., $\theta:V(G) \longrightarrow [0,1]$ and $\mathcal{P}:E(G) \longrightarrow (0,1]$. In \textit{information diffusion}, vertex and edge weights are called node threshold and diffusion probability, respectively \cite{gruhl2004information}. More the value of $\theta_i$, more hard to influence the user $u_i$ and more the value of $p_{ij}$, it is more probable that $u_i$ can influence $u_j$.  For any user $u_i \in V(G)$, its \textit{incoming neighbors} and \textit{outgoing neighbors} $\mathcal{N}^{in}(u_i)$ and $\mathcal{N}^{out}(u_i)$ are defined as: $\mathcal{N}^{in}(u_i) = \left\{ u_j \vert (u_ju_i) \in E(G) \right\}$ and $\mathcal{N}^{out}(u_i) = \left\{ u_j \vert (u_iu_j) \in E(G) \right\}$, respectively. For any user $u_i \in V(G)$, its \textit{indegree} and \textit{outdegree} is defined as $deg^{in}(u_i)= \vert \mathcal{N}^{in}(u_i) \vert$ and $deg^{out}(u_i)= \vert \mathcal{N}^{out}(u_i) \vert$, respectively. A \textit{path} in a directed graph is a sequence of vertices without repetition, such that between every consecutive vertices there will be an \textit{edge}. Two users are connected in the graph $G$, if there exists a directed path between them. A directed graph is said to be connected, if there exists a path between every pair of users.
\subsection{Relation between Target Set Selection and Other Graph Theoretic Problems}
\par The TSS Problem is a more generalized version of many standard graph theoretic problems discussed and mentioned in the literature, such as \textit{dominating set with threshold} \cite{harant1999dominating}, \textit{vector domination problem} \cite{raman2008parameterized}, \textit{k\mbox{-}tuple dominating set} \cite{klasing2004hardness} (in all these problems instead of multiple rounds, diffusion can run only for one round), \textit{vertex cover} \cite{chen2009approximability} (in this problem, vertex threshold is set equal to the number of neighbors of the node), \textit{irreversible k\mbox{-}conversion problem} \cite{dreyer2009irreversible}, \textit{r\mbox{-}neighbor bootstrap percolation problem} \cite{balogh2010bootstrap} (where the threshold of each vertex is $k$ or $r$ respectively) and \textit{dynamic monopolies} \cite{peleg2002local} (in this case, threshold is half of the neighbors of the user). 
\subsection{Approximation Algorithm} Most of the \textit{optimization problems} arising in real life are NP\mbox{-}Hard \cite{garey2002computers}. Hence, we cannot expect to solve them by any deterministic algorithm in polynomial time. So, the goal is to get an approximate solution  of the problem within affordable time. Approximation algorithms serve this purpose and also provide the worst case guarantee on solution quality. For a maximization problem $\mathcal{P}$, let $\mathcal{A}$ be an algorithm, which provides its solution and $\mathcal{I}$ be the set of all possible input instances of $\mathcal{P}$. For an input instance $I$ of $\mathcal{P}$; let, $\mathcal{A}^{*}(I)$ is the optimal solution and $\mathcal{A}(I)$ is the solution generated by the algorithm $\mathcal{A}$. Now, $\mathcal{A}$ will be called an $\alpha$\mbox{-}factor \textit{absolute approximation algorithm}, if $\forall I \in \mathcal{I}$, $\vert \mathcal{A}^{*}(I)- \mathcal{A}(I) \vert \leq \alpha$ and $\alpha$\mbox{-}factor \textit{relative approximation algorithm}, if $\forall I \in \mathcal{I}$, $max\{\frac{\mathcal{A}^{*}(I)}{\mathcal{A}(I)}, \frac{\mathcal{A}(I)}{\mathcal{A}^{*}(I)} \} \leq \alpha$ ($\mathcal{A}(I),\mathcal{A}^{*}(I) \neq 0$) \cite{williamson2011design}. Section \ref{Sec:AAPG} of this paper describes relative approximation algorithms for solving SIM Problem.
\subsection{Parameterized Complexity Theory}
  Parameterized complexity theory is another way of dealing with NP\mbox{-}Hard optimization problems. It aims to classify computational problems based on the inherent difficulty with respect to multiple parameters related to the problem. There are several \textit{complexity classes} in parameterized complexity theory. The class FPT (\textit{Fixed Parameter Tractable}) contains the problems for which, any problem with instances $(x,k) \in \mathcal{I}$, where $x$ is the input , $k$ is the parameter and $ \mathcal{I}$ is the set of instances; its running time will be of $\mathcal{O}(f(k) \vert x \vert ^{\mathcal{O}(1)})$, where $f(k)$ is the function depending on only $k$ and $\vert x \vert$ denotes the length of the input. $W$ hierarchy is the collection of complexity classes with the property $W[0]=FPT$ and $W[i] \subseteq W[j]$ $\forall i \leq j$ \cite{downey1998parameterized}. Many normal computational problems occupy the lower levels of hierarchy, i.e., $W[1]$ and $W[2]$. In Section \ref{Sec:Hard}, we have described hardness results of TSS Problem in parameterized complexity theoretic setting. 
\subsection{Information Diffusion in a Social Network} \label{BID}
 Diffusion phenomena in a networked system has got attention from different disciplines, such as \textit{epidemiology} (how diseases spread in a human contact network?) \cite{salathe2010high}, \textit{social network analysis} (how information propagates in a social network?) \cite{xu2010information}, \textit{computer network} (how computer virus propagates in an e\mbox{-}mail network?) \cite{zou2007modeling} etc. \textit{Information Diffusion} in an on\mbox{-}line social networks is a phenomenon by which word-of-mouth effect occurs electronically. Hence, the mechanism of information diffusion is very well studied \cite{kimura2006tractable} \cite{valente1995network}. To study the diffusion process, there are  some models  in the literature \cite{heidari2016modeling}. Nature of these models varies from \textit{deterministic} to \textit{probabilistic}. Here, we have described some well studied \textit{information diffusion models} from the literature.
\begin{itemize}
	\item \textit{Independent Cascade Model} (IC Model) \cite{shakarian2015independent}: This is one of the well studied  probabilistic diffusion models used by Kempe et al. \cite{kempe2003maximizing} in their seminal work of \textit{social influence maximization}. In this model, a node can either be in active  state (i.e., influenced) or in inactive state (i.e., not influenced). Initially (i.e., at $t=0$), all the nodes except the seeds are inactive.  Every active node (say, $u_i$) at time stamp $t$ will get a chance to activate its currently \textit{inactive}  neighbor ($u_j \in \mathcal{N}^{out}(u_i)$ and $u_j$ is inactive) with probability as their edge weight. If $u_i$ succeeds, then $u_j$ will become an active node in time stamp $t+1$. A node can change its state from inactive to active but not from active to inactive. This cascading process will be continued until no more active node is there in a time stamp. Suppose, this diffusion process starts at $t=0$ and continued till $t=\mathcal{T}$ and $\mathcal{A}_{t}$ denotes the set of active nodes till time stamp $t$, where $t \in [0, \mathcal{T}]$, then
	\begin{center}
		$\mathcal{A}_{0} \subseteq \mathcal{A}_{1} \subseteq \dots \subseteq \mathcal{A}_{t} \subseteq \mathcal{A}_{t+1} \subseteq \dots \subseteq \mathcal{A}_{\mathcal{T}} \subseteq V(\mathcal{G})$.
	\end{center}
	Node $u_i$ is said to be active at time stamp $t$, if $u_i \in \mathcal{A}_{t} \setminus \mathcal{A}_{t-1}$.
	\item \textit{Linear Threshold Model} (\textit{LT Model}) \cite{shakarian2015independent}: This is another probabilistic diffusion model proposed by Kempe et al. \cite{kempe2003maximizing}. In this model, for any node (say $u_i$), all its neighbors who are activated just at previous time stamp together make a try to activate that node. This activation process will be successful, if the sum of the incoming active neighbor's probability becomes either greater than or equal to the node's threshold, i.e., $\forall u_j \in \mathcal{N}^{in}(u_i)$,  if $\sum_{\forall u_j \in \mathcal{N}^{in}(u_i); u_j \in \mathcal{A}_{t}} p_{ji} \geq \theta_i$ then, $u_i$ will become active at time stamp $t+1$. This method will be continued until no more activation is possible. In this model, we can use the negative influence, which is not possible in IC Model. Later, several extensions of this two fundamental models have been proposed \cite{yang2010modeling}.\\
	In both IC as well as LT Model, it is assumed that diffusion probability between two users is known. However, later there were several studies for computing diffusion probability  \cite{saito2011learning} \cite{saito2008prediction} \cite{goyal2010learning} \cite{saito2010selecting} \cite{kimura2009finding}.
	\item \textit{Shortest Path Model} (\textit{SP Model}): This is a special case of IC Model proposed by Kimura et al. \cite{kimura2006tractable}. In this model, an inactive node will get a chance to become active only through the shortest path from the initially active nodes, i.e., at $t=\underset{u \in \mathcal{A}_{0},v \in V(G)\setminus \mathcal{A}_{0}} {min} dist(u,v)$. A slightly different variation of SP Model proposed by the same author is \textit{SP1 Model}, which tells that an inactive node will get a chance of activation at $t=\underset{u \in \mathcal{A}_{0},v \in V(G)\setminus \mathcal{A}_{0}} {min} dist(u,v)$ and $t=\underset{u \in \mathcal{A}_{0},v \in V(G)\setminus \mathcal{A}_{0}} {min} dist(u,v)+1$.
	\item \textit{Majority Threshold Model} (\textit{MT Model}): This is the deterministic threshold model proposed by Valente \cite{valente1996social}. In this model, the vertex threshold is defined as $\theta_i=\ceil*{\frac{deg(u_i)}{2}}$, which means that a node will become active, when atleast half of its neighbors are already active in nature. 
	\item \textit{Constant Threshold Model} (\textit{CT Model}): This is another deterministic diffusion model, where vertex threshold can be any value from 1 to its degree, i.e., $\theta_i \in [deg(u_i)]$.
	\item \textit{Unanimous Threshold Model} (\textit{UT model}) \cite{chen2009approximability}: This is the most influence resistant model of diffusion. In this model, for each node in the network, its threshold value is set to its degree i.e., $\forall u_i \in V(G)$, $\theta_i=deg(u_i)$.
\end{itemize} 
  There are many other diffusion models, such as \textit{weighted cascade model}, where edge weight will be the reciprocal of the degree of the node; \textit{trivalency model}, where the edge weights are uniformly taken from the set: $\{0.1, 0.01, 0.001 \}$ etc. Readers require a detailed and exhaustive treatment on information diffusion models may refer to \cite{zhang2014recent}. 
\section{SIM Problem and its Variants}  \label{Sec:VTSSP}
In literature, SIM problem has been studied since early two thousand. Initially, this problem was introduced by Domingos and Richardson in the context of viral marketing \cite{domingos2001mining}. Due to its substantial practical importance across multiple domains, different variants of this problem have been introduced. In this section, we will describe them one by one.
\paragraph{Basic SIM Problem \cite{ackerman2010combinatorial}:} In the basic version of the \textit{TSS Problem} along with a \textit{directed social network} $G(V, E, \theta, \mathcal{P})$, we are given two integers: $k$ and $\lambda$, and asked to find out a subset of atmost $k$ nodes such that after the diffusion process is over atleast $\lambda$ number of nodes are activated. Mathematically, this problem can be stated as follows:
\begin{mdframed}[style=MyFrame]
	\begin{center}
		\textit{\textbf{Instance:} A Directed Graph $G(V, E, \theta, \mathcal{P})$, $\lambda \in [n]$ and $k \in \mathbb{Z}^{+}$.}\\
		\textit{\textbf{Problem:}Basic TSS Problem [Find out a $\mathcal{S} \subset V(G)$, such that $\vert \mathcal{S} \vert \leq k$, and $\vert \sigma(\mathcal{S}) \vert \geq \lambda$].}\\
		\textit{\textbf{Output:} The Seed Set for Diffusion $\mathcal{S} \subset V(G)$ and $\vert \mathcal{S} \vert \leq k$.}\\
	\end{center}
\end{mdframed}
\paragraph{Top k-node Problem / Social Influence Maximization Problem (SIM Problem) \cite{narayanam2011shapley}:} This variant of the problem is most well studied. For a given social network $G(V, E, \theta, \mathcal{P})$, this problem asks to choose a set $\mathcal{S}$ of $k$ nodes (i.e., $\mathcal{S} \subset V(G)$ and $\vert \mathcal{S} \vert=k$) such that the maximum number of nodes of the network become influenced at the end of diffusion process, i.e., $\sigma(\mathcal{S})$ will be maximized. Most of the algorithms presented in Section \ref{Sec:SolTSS} are solely develop for solving this problem. Mathematically, the \textit{Problem of Top k-node Selection} will be like the following:
\pagebreak
\begin{mdframed}[style=MyFrame]
	\begin{center}
		\textit{\textbf{Instance:} A Directed Graph $G(V, E, \theta, \mathcal{P})$ and $k \in \mathbb{Z}^{+}$.}\\
		\textit{\textbf{Problem:}Top k-node Problem [Find out a $\mathcal{S} \subset V(G)$ where $\vert \mathcal{S} \vert=k$ such that and for any other $\mathcal{S}^{'} \subset V(G)$ with $\vert \mathcal{S}^{'} \vert=k$, $\sigma(\mathcal{S}) \geq \sigma(\mathcal{S}^{'})$].}\\
		\textit{\textbf{Output:} The Seed Set for Diffusion $\mathcal{S} \subset V(G)$ and $\vert \mathcal{S} \vert = k$.}\\
	\end{center}
\end{mdframed}
\paragraph{Influence Spectrum Problem} \cite{nguyen2017social} In this problem, along with the social network $G(V, E, \theta, \mathcal{P})$, we are also given with two integers: $k_{lower}$ and $k_{upper}$ with $k_{upper} > k_{lower}$. Our goal is to choose a set $\mathcal{S}$ for each $k \in [k_{lower}, k_{upper}]$, such that social influence in the network ($\sigma(\mathcal{S})$) is maximum in each case. Intutively, solving one instance of this problem is equivalent to solving $(k_{upper} - k_{lower} + 1)$ instances of SIM problem. As viral marketing is basically done in different phases and in each phase, seed set of different cardinalities can be used, influence spectrum problem appears in a natural way. Mathematically, influence spectrum problem can be written as follows:
\begin{mdframed}[style=MyFrame]
	\begin{center}
		\textit{\textbf{Instance:} A Directed Graph $G(V, E, \theta, \mathcal{P})$ and $k_{lower},k_{upper} \in \mathbb{Z}^{+}$ with $k_{upper} > k_{lower}$.}\\
		\textit{\textbf{Problem:}Influence Spectrum Problem [Find out a $\mathcal{S} \subset V(G)$ with $\vert \mathcal{S} \vert=k$, $\forall k \in [k_{lower}, k_{upper}]$ such that and for any other $\mathcal{S}^{'} \subset V(G)$ with $\vert \mathcal{S}^{'} \vert=k$, $\sigma(\mathcal{S}) \geq \sigma(\mathcal{S}^{'})$].}\\
		\textit{\textbf{Output:} The Seed Set for Diffusion $\mathcal{S} \subset V(G)$ and $\vert \mathcal{S} \vert = k$ for each $k \in [k_{lower}, k_{upper}]$.}\\
	\end{center}
\end{mdframed}
\paragraph{$\lambda$ Coverage Problem \cite{narayanam2011shapley}:} This is another variant of SIM Problem, which considers the minimum number of influenced nodes required at the end of diffusion. For a given social network  $G(V, E, \theta, \mathcal{P})$ and a constant $\lambda \in [n]$, this problem asks to find a subset $\mathcal{S}$ of its nodes with minimum cardinality, such that at least $\lambda$ number of nodes will be influenced at the end of diffusion process. Mathematically, this problem can be described in the following way:
\begin{mdframed}[style=MyFrame]
	\begin{center}
		\textit{\textbf{Instance:} A Directed Graph $G(V, E, \theta, \mathcal{P})$ and $\lambda \in [n]$.}\\
		\textit{\textbf{Problem:} $\lambda$ Coverage Problem [Find out the most minimum cardinality subset $\mathcal{S} \subset V(G)$ such that  $\vert \sigma(\mathcal{S}) \vert \geq \lambda$ ].}\\
		\textit{\textbf{Output:} The minimum cardinality seed set $\mathcal{S}$ for diffusion.}\\
	\end{center}
\end{mdframed}
\paragraph{Weighted Target Set Selection Problem (WTSS Problem) \cite{raghavan2015weighted}:} This is another (infect weighted) variant of SIM Problem. Along with a social network $G(V, E, \theta, \mathcal{P})$, we are given another \textit{vertex weight function}, $\phi :V(G) \rightarrow \mathbb{N}_0$, signifying the cost associated with each vertex. This problem asks to find out a subset $\mathcal{S}$, which \textit{minimizes} total \textit{selection cost}, and also all the nodes will be influenced at the end of diffusion. Mathematically, this problem can be stated as follows:
\begin{mdframed}[style=MyFrame]
	\begin{center}
		\textit{\textbf{Instance:} A Directed Graph $G(V, E, \theta, \mathcal{P})$, vertex cost function $\phi :V(G) \rightarrow \mathbb{N}_0$.}\\
		\textit{\textbf{Problem:} Weighted TSS Problem [Find out the subset $\mathcal{S} \subset V(G)$ such that $\phi(\mathcal{S})$ is minimum and $\vert \sigma(\mathcal{S}) \vert = n$].}\\
		\textit{\textbf{Output:} The Seed Set for Diffusion $\mathcal{S} \subset V(G)$ with minimum $\phi(\mathcal{S})$ value.}\\
	\end{center}
\end{mdframed}
\paragraph{r\mbox{-}round min\mbox{-}TSS Problem \cite{charikar2016approximating}:} It is a variant of SIM Problem, which considers the number of rounds required to complete the diffusion process. Along with a \textit{directed graph} $G(V, E, \theta, \mathcal{P})$, we are given the maximum number of allowable rounds $r \in \mathbb{Z}^{+}$, and asks to find out a minimum cardinality seed set $\mathcal{S}$, which activates all the nodes of the network within $r$\mbox{-}round. Mathematically, this problem can be described as follows:
\begin{mdframed}[style=MyFrame]
	\begin{center}
		\textit{\textbf{Instance:} A Directed Graph $G(V, E, \theta, \mathcal{P})$ and $r \in \mathbb{Z}^{+}$.}\\
		\textit{\textbf{Problem:} r\mbox{-}round min\mbox{-}TSS Problem [Find out the most minimum cardinality subset $\mathcal{S}$ such that $\cup_{i=1}^{r}\sigma_i(\mathcal{S})=V(G)$].}\\
		\textit{\textbf{Output:} The Seed Set for Diffusion $\mathcal{S} \subset V(G)$.}\\
	\end{center}
\end{mdframed}
Here, $\sigma_i(\mathcal{S})$ denotes the set of influenced nodes from the seed set $\mathcal{S}$ at the $i$\mbox{-}th round of diffusion.
\paragraph{Budgeted Influence Maximization Problem (BIM Problem) \cite{nguyen2013budgeted}:} This is another variant of SIM Problem, which is recently gaining popularity. Along with a \textit{directed graph} $G(V, E, \theta, \mathcal{P})$, we are given with a cost function $\mathcal{C}: V(G) \longrightarrow \mathbb{Z}^{+}$ and a fixed budget $\mathcal{B} \in \mathbb{Z}^{+}$. Cost function $\mathcal{C}$ assigns a nonuniform selection cost to every vertex of the network, which is the amount of incentive need to be paid, if that vertex is selected as a seed node. This problem asks for selecting a seed set within the budget, which maximizes the spread of influence in the network.
\begin{mdframed}[style=MyFrame]
	\begin{center}
		\textit{\textbf{Instance:} A Directed Graph $G(V, E, \theta, \mathcal{P})$, a cost function $\mathcal{C}: V(G) \longrightarrow \mathbb{Z}^{+}$ and affordable budget $\mathcal{B} \in \mathbb{Z}^{+}$.}\\
		\textit{\textbf{Problem:} Budgeted Influence Maximization Problem [Find out the seed set ($\mathcal{S}$) such that $\underset{u \in \mathcal{S}}{\sum} \mathcal{C}(u) \leq \mathcal{B}$ and for any other seed set $\mathcal{S}^{'}$ with $\underset{v \in \mathcal{S}^{'}}{\sum} \mathcal{C}(v) \leq \mathcal{B}$, $\vert \sigma(\mathcal{S}) \vert \geq \vert \sigma(\mathcal{S}^{'} \vert$)].}\\
		\textit{\textbf{Output:} The Seed Set for Diffusion $\mathcal{S} \subset V(G)$ with $\underset{u \in \mathcal{S}}{\sum} \mathcal{C}(u) \leq \mathcal{B}$.}\\
	\end{center}
\end{mdframed}
\paragraph{$(\lambda,\beta, \alpha)$ TSS Problem \cite{cicalese2014latency}:} This is another variant of TSS Problem, which considers the maximum cardinality of the seed set ($\beta$), maximum allowable diffusion rounds ($\lambda$), and number of influenced nodes at the end of diffusion process ($\alpha$) all together. Along with the input graph $G(V, E, \theta, \mathcal{P})$, we are given with the parameters $\lambda, \beta$ and $\alpha $. Mathematically, this problem can be stated as follows:
\begin{mdframed}[style=MyFrame]
	\begin{center}
		\textit{\textbf{Instance:} A Directed Graph $G(V, E, \theta, \mathcal{P})$, three parameters $\lambda, \beta  \in \mathbb{N}$ and $\alpha \in [n]$.}\\
		\textit{\textbf{Problem:}$(\lambda,\beta, \alpha)$ TSS Problem [Find out the subset $\mathcal{S} \subset V(G)$ such that $\vert \mathcal{S} \vert \leq \beta$, $ \vert \cup_{i=1}^{\lambda}\sigma_i(\mathcal{S}) \vert \geq \alpha$].}\\
		\textit{\textbf{Output:} The Seed Set for Diffusion $\mathcal{S} \subset V(G)$ and $\vert \mathcal{S} \vert \leq \beta$.}\\
	\end{center}
\end{mdframed} 
\paragraph{$(\lambda,\beta, A)$ TSS Problem \cite{cicalese2014latency}:} This is a slightly different from the $(\lambda,\beta, \alpha)$ TSS problem, in which instead of the required number of the nodes after the diffusion process, it explicitly maintains which nodes should be influenced. Along with the input social network $G(V, E, \theta, \mathcal{P})$, we are also given with maximum allowable rounds ($\lambda$), maximum cardinality of the seed set ($\beta$), and set of nodes $A \subseteq V(G)$ need to be influenced at the end of diffusion process as input. This problem asks for selecting a seed set of maximum $\beta$ elements, which will influence all the nodes in $A$ within $\lambda$ rounds of diffusion. Mathematically, the problem can be stated as follows: 
\begin{mdframed}[style=MyFrame]
	\begin{center}
		\textit{\textbf{Instance:} A Directed Graph $G(V, E, \theta, \mathcal{P})$, $A \subseteq V(G)$ and two parameters $\lambda, \beta \in \mathbb{N}$.}\\
		\textit{\textbf{Problem:}$(\lambda,\beta, A)$ TSS Problem [Find out the subset $\mathcal{S} \subset V(G)$ such that $\vert \mathcal{S} \vert \leq \beta$, $A \subseteq \cup_{i=1}^{\lambda}\sigma_i(\mathcal{S})$].}\\
		\textit{\textbf{Output:} The Seed Set for Diffusion $\mathcal{S} \subset V(G)$ and $\vert \mathcal{S} \vert \leq \beta$.}\\
	\end{center}
\end{mdframed}
\paragraph{$(\lambda, A)$ TSS Problem \cite{cicalese2014latency}:} This is slightly different from $(\lambda,\beta, A)$ TSS Problem. Here, we are interested in finding the minimum cardinality seed set, such that within some fixed numbers of diffusion rounds ($\lambda$), a subset of the nodes ($A$) will be influenced. Mathematically, the problem can be stated as follows:
\begin{mdframed}[style=MyFrame]
	\begin{center}
		\textit{\textbf{Instance:} A Directed Graph $G(V, E, \theta, \mathcal{P})$, $A \subset V(G)$ and  $\lambda \in \mathbb{N}$.}\\
		\textit{\textbf{Problem:}$(\lambda, A)$ TSS Problem [Find out the subset $\mathcal{S}$ such that $A \subseteq \cup_{i=1}^{\lambda}\sigma_i(\mathcal{S})$ and for any other $\mathcal{S}^{'}$ with $\vert \mathcal{S}^{'} \vert < \vert \mathcal{S} \vert$ $A \not \subseteq \cup_{i=1}^{\lambda}\sigma_i(\mathcal{S}^{'})$].}\\
		\textit{\textbf{Output:} Minimum cardinality Seed Set for Diffusion $\mathcal{S} \subset V(G)$.}\\
	\end{center}
\end{mdframed}
\par We have described different variants of TSS Problem in social networks available in the literature. It is surprising to see that only Top\mbox{-}k node Problem has been studied, in depth. 
\section{Hardness Results of TSS Problem} \label{Sec:Hard}
In this section, we have described hardness results of SIM Problem under both general as well as parameterized complexity theoretic perspective. Initially, the problem of social influence maximization was posed by Domingos and Richardson \cite{domingos2001mining} \cite{richardson2002mining} in the context of viral marketing. However, Kempe et al. \cite{kempe2003maximizing} was the first to investigate the computational issues of the problem. They were able to show that  SIM Problem under IC and LT Model is a special case of \textit{Set Cover Problem} and \textit{Vertex Cover Problem}, respectively. Both the  set cover and vertex cover problems are well-known \textit{NP\mbox{-}Hard} problems \cite{garey2002computers}. The conclusion is presented as Theorem \ref{Th:1}. 
\begin{mythm} \label{Th:1} \cite{kempe2003maximizing}
	Social Influence Maximization Problem is NP-Hard for both  IC as well as LT model and also NP-Hard to approximate within a factor of $n^{(1-\epsilon)} \ \forall \epsilon>0$.
\end{mythm}
  Chen \cite {chen2009approximability} studied variant of SIM Problem namely \textit{$\lambda$ Coverage Problem}. His study was different from Kempe et al.'s \cite{kempe2003maximizing} study in two ways. First one is, Kempe et al. \cite{kempe2003maximizing} investigated the Top\mbox{-}$k$ node problem, whereas Chen \cite {chen2009approximability} studied the $\lambda$\mbox{-}coverage problem. Secondly, Kempe et al. \cite{kempe2003maximizing} studied  the diffusion process under IC and LT Models, which are probabilistic in nature, whereas Chen \cite {chen2009approximability} considered all the \textit{deterministic diffusion models} like \textit{majority threshold model}, \textit{constant threshold model} and \textit{unanimous threshold model}. In general,  for the $\lambda$ Coverage Problem, Chen \cite {chen2009approximability} came up with a seminal result presented in Theorem \ref{Th:2}.
\begin{mythm} \cite{chen2009approximability} \label{Th:2}
 	TSS Problem cannot be approximated with in the constant factor $\mathcal{O}(2^{\log^{(1-\epsilon)}n})$ unless $NP \subset DTIME(n^{polylog (n)})$ for any fixed constant $\epsilon > 0$. 
 \end{mythm}
\par This theorem can be proved by a reduction from the \textit{Minimum Representative Problem} given in \cite{kortsarz2001hardness}. Next, they have shown that in \textit{majority threshold model} also, \textit{$\lambda$\mbox{-}coverage problem} follows the similar result as presented in Theorem \ref{Th:2}. However,  when $\theta(u)=1$, $\forall u \in V(G)$ then TSS Problem can be solved very intuitively as targeting one node in each component results into the activation of all the nodes of the network. Surprisingly, this problem becomes hard, when we allow the vertex threshold to be at most 2, i.e., $\theta(u) \leq 2$ $\forall u \in V(G)$. They proved the following result in this regard.
 \begin{mythm} \cite{chen2009approximability}
 	The TSS Problem is NP\mbox{-}Hard, when thresholds are at most 2, even for bounded bipartite graphs.
 \end{mythm}
\par This theorem can be proved by a reduction from a variant of 3\mbox{-}SAT Problem presented in \cite{tovey1984simplified}. Moreover, Chen \cite{chen2009approximability} has shown that for \textit{unanimous threshold model}, the \textit{TSS Problem} is equivalent to \textit{vertex cover problem}, which is a well-known NP\mbox{-}Complete Problem.
 \begin{mythm} \cite{chen2009approximability}
 	If all the vertex thresholds of the graph are unanimous (i.e. $\forall u \in V(G)$, $\theta(u)=deg(u)$), then the TSS Problem is identical to vertex cover problem.
  \end{mythm}	
\par  Chen \cite{chen2009approximability} has also shown that if the underline graph is tree, then the TSS Problem can be solved in polynomial time and they have also given the \textit{ALG\mbox{-}Tree} Algorithm, which does this computation. To the best of the authors' knowledge, there is no other literature, which focuses on the hardness analysis of the TSS Problem in traditional complexity theoretic perspective. We have summarized the results in Table \ref{Tab:TC}.
\par Now, we describe the hardness results based on the parameterized complexity theoretic perspective. For basic notions about \textit{parameterized complexity}, readers may refer to \cite{downey2013fundamentals}. Bazgan et al. \cite{bazgan2014parameterized} showed that SIM Problem under constant threshold model (CTM) does not have any parameterized approximation algorithm with respect to the parameter \textit{seed set size}. Chopin et al. \cite{chopin2014constant}, \cite{Chopin2012} studied the TSS Problem in parameterized settings with respect to the  parameters related to network cohesiveness like \textit{clique cover number} (number of cliques required to cover all the vertices of the network \cite{karp1972reducibility}), \textit{distance to clique} (number of vertices need to be deleted to obtain a clique), \textit{cluster vertex deletion number} (number of vertices to delete in order to obtain a collection of disjoint cliques); parameters related to network density like \textit{distance to cograph}, \textit{distance to interval graph}; parameters related to sparsity of the network, namely \textit{vertex cover number} (number of vertices to remove to obtain an edgeless graph), \textit{feedback edge set number and feedback vertex set number} (number of edges or vertices to remove to obtain a forest), \textit{pathwidth}, \textit{bandwidth}. It is interesting to note that computing all the parameters except \textit{feedback edge set number} is NP\mbox{-}Hard problem. The version of TSS Problem, they have worked with is $\lambda$\mbox{-}coverage problem with $\lambda=n$. They came up with the following two important results related to the sparsity parameters of the network:
 \begin{mythm} \cite{chopin2014constant}
 	TSS Problem with majority threshold model is W[1] hard even
 	with respect to the combined parameter feedback vertex set, distance to co-graph, distance to interval graph, and path width.
 \end{mythm}
 \begin{mythm} \cite{chopin2014constant}
 	TSS Problem is fixed-parameter tractable with respect to the parameter bandwidth.
\end{mythm}
For proving the above two theorems, authors have used reduction rules used in \cite{nichterlein2013tractable} and \cite{nichterlein2010tractable}. Results related to dense structure property of the network is given in Theorems \ref{Th:7} through \ref{Th:9}.
\begin{mythm} \label{Th:7}
TSS Problem is W[1]\mbox{-}Hard with parameter cluster vertex deletion number.
\end{mythm}
\begin{mythm}
TSS Problem is NP\mbox{-}Hard and W[2] Hard with respect to the parameter target set size ($k$), even on graphs with clique cover number of two.
\end{mythm}
\begin{mythm}  \label{Th:9}
	TSS Problem is fixed parameter tractable with respect to the parameter `distance l to clique', if the threshold function satisfies following properties $\theta(u)>g(l) \Rightarrow \theta(u)=f(\Gamma(u))$ $\forall u \in V(G)$, $f:P(V(G)) \longrightarrow \mathbb{N}$ and $g: \mathbb{N} \longrightarrow \mathbb{N}$.
\end{mythm} 
For detailed proof of Theorems \ref{Th:7} through \ref{Th:9}, readers may refer to \cite{chopin2014constant}. All the results related to the parameterized complexity theory has been summarized in Table \ref{Tab:PC}.

\begin{table} 
\begin{center}
    \begin{tabular}{ | p{3 cm} | p{2.2 cm} | p{6.8 cm} |}
    \hline
    \textbf{Name of the Problem} & \textbf{Diffusion Model} &  \textbf{Major Findings} \\
     \hline
    \multirow{4}{*}{SIM} & IC Model & A special case of set cover problem and hence NP\mbox{-}Hard. \\
    \cline{2-3}
     & LT Model & A special case of vertex cover problem and hence NP\mbox{-}Hard.\\
    \hline
    \multirow{13}{*}{$\lambda$\mbox{-}Coverage Problem} & MT Model & Not only NP\mbox{-}Hard as well as can not be approximated in the constant factor $\mathcal{O}(2^{\log^{(1-\epsilon)}n})$ unless $NP \subset DTIME(n^{polylog (n)})$\\
     \cline{2-3}
    & CT Model with $\theta(u)=1$, $\forall u \in V(G)$ & Can be solved trivially by selecting a vertex from each component of the network.\\
   \cline{2-3}
   & CT Model with $\theta(u) \leq 2$, $\forall u \in V(G)$ & NP\mbox{-}Hard even for bounded bipartite graphs.\\
  \cline{2-3}
   & UT Model & Identical to vertex cover problem and hence NP\mbox{-}Hard\\
     \hline
    \end{tabular}
\end{center}
\caption{Hardness results of TSS Problem and its variants in traditional complexity theory perspective.}
\label{Tab:TC}
\end{table}

\begin{table}
\begin{center}
    \begin{tabular}{ | p{2.2 cm} | p{2.5 cm} | p{3 cm} | p{4cm} |}
    \hline
    \textbf{Name of the Problem} & \textbf{Diffusion Model} & \textbf{Parameter} & \textbf{Major Findings} \\ 
    \hline
     SIM & CT Model with $\theta(u) \in [deg(u)]$  & Seed Set Size & Does not have any parameterized approximation algorithm. \\
     \hline
     $\lambda$\mbox{-}coverage Problem with $\lambda=n$ & MT Model & Feedback vertex set number, Pathwidth, Distance to cograph, Distance to interval graph & The problem is $W[1]$\mbox{-}Hard. \\ 
     \hline 
    $\lambda$\mbox{-}coverage Problem with $\lambda=n$ & GT Model & Cluster vertex deletion number & The problem is $W[1]$\mbox{-}Hard \\ 
    \hline
    $\lambda$\mbox{-}coverage Problem with $\lambda=n$ & CT Model & Cluster vertex deletion number & The problem is fixed parameter tractable. \\
    \hline
  $\lambda$\mbox{-}coverage Problem with $\lambda=n$ & GT Model & Seed set size &  The problem is $W[2]$\mbox{-}Hard \\  
    \hline
$\lambda$\mbox{-}coverage Problem with $\lambda=n$ & MT Model, CT Model & distance to clique & The problem is fixed parameter tractable.\\
    \hline
    \end{tabular}
\end{center}
\caption{Hardness results of TSS Problem and its variants in parameterized complexity theory perspective.}
\label{Tab:PC}
\end{table}
\section{Major Research Challenges} \label{Sec:MRC}
Before entering into the critical review of the existing solution methodologies, in this section, we provide a brief discussion on major research challenges concerned with the SIM Problem. This will help the reader to understand which category of solution methodology can handle what challenge. 
\begin{itemize}
\item \textbf{Trade of Between Accuracy and Computational Time:} From the discussion in Section \ref{Sec:Hard}, it is now well understood that the SIM Problem is computationally hard from both traditional as well as parameterized complexity theoretic prospective, in general. Hence, for some given $k \in \mathbb{Z}^{+}$, obtaining the most influential $k$ nodes within feasible time is not possible. In this scenario, the intuitive approach could be to use some heuristic method for selecting seed nodes. This will lead to less time for seed set generation. However, the number of influenced nodes generated by the seed nodes could be also arbitrarily less. In this situation, it is an important issue to design algorithms, which will run in affordable time and also, the gap between the optimal spread and the spread due to the seed set selected by an algorithm will be as much less as possible.  
\item \textbf{Breaking the Barrier of Submodularity:} In general, the social influence function $\sigma(.)$ is submodular (Discussed in Section \ref{Sec:AAPG}). However, in many practical situations, such as \textit{opinion and topic specific influence maximization}, the social influence function may not be submodular \cite{li2013influence} \cite{gionis2013opinion}. This happens because one node can switch its state from positive opinion to negative opinion and the vice\mbox{-}versa. In this scenario, solving the SIM Problem may be more challenging due to the absence of submodularity property in the social influence function.
\item \textbf{Practicality of the Problem:} In general, the 
SIM Problem takes many assumptions, such as every selected seed will perform up to expectation in the spreading process, influencing each node of the network is equally important etc. This assumptions may be unrealistic in some situations. Assume the case of \textit{target advertisement}, where instead of all the nodes, a set of target nodes are chosen and the aim is to maximize the influence within the target nodes \cite{epasto2017real} \cite{ke2018finding}. In another way, due to the probabilistic nature of diffusion, a seed node may not perform up to expectation in the influence spreading process. Solving the SIM Problem and its variants will be more challenging, if we relax these assumptions.
\item \textbf{Scalability:} Real life social networks have millions of nodes and billions of edges. So, solving the SIM and related problems for real life social networks, scalability should be an important issue for any solution methodology. 
\item \textbf{Theoretical Challenges:} For a computational problem, any of its solution methodology is concerned with two aspects. First one is the \textit{computational time}. This is measured as the execution time, when the methodology is implemented with real life problem instances. The second one is the \textit{computational complexity}. This is measured as the \textit {asymptotic bound} of the methodology. Theoretical research on any computational problem always concerned with the second aspect of the problem. Hence, the theoretical challenge for the SIM Problem is to design algorithms with good asymptotic bounds.  
\end{itemize}
\section{Solutions Methodologies} \label{Sec:SolTSS} 
\begin{figure}
 \centering
  \includegraphics[height=8 cm, width= 15 cm]{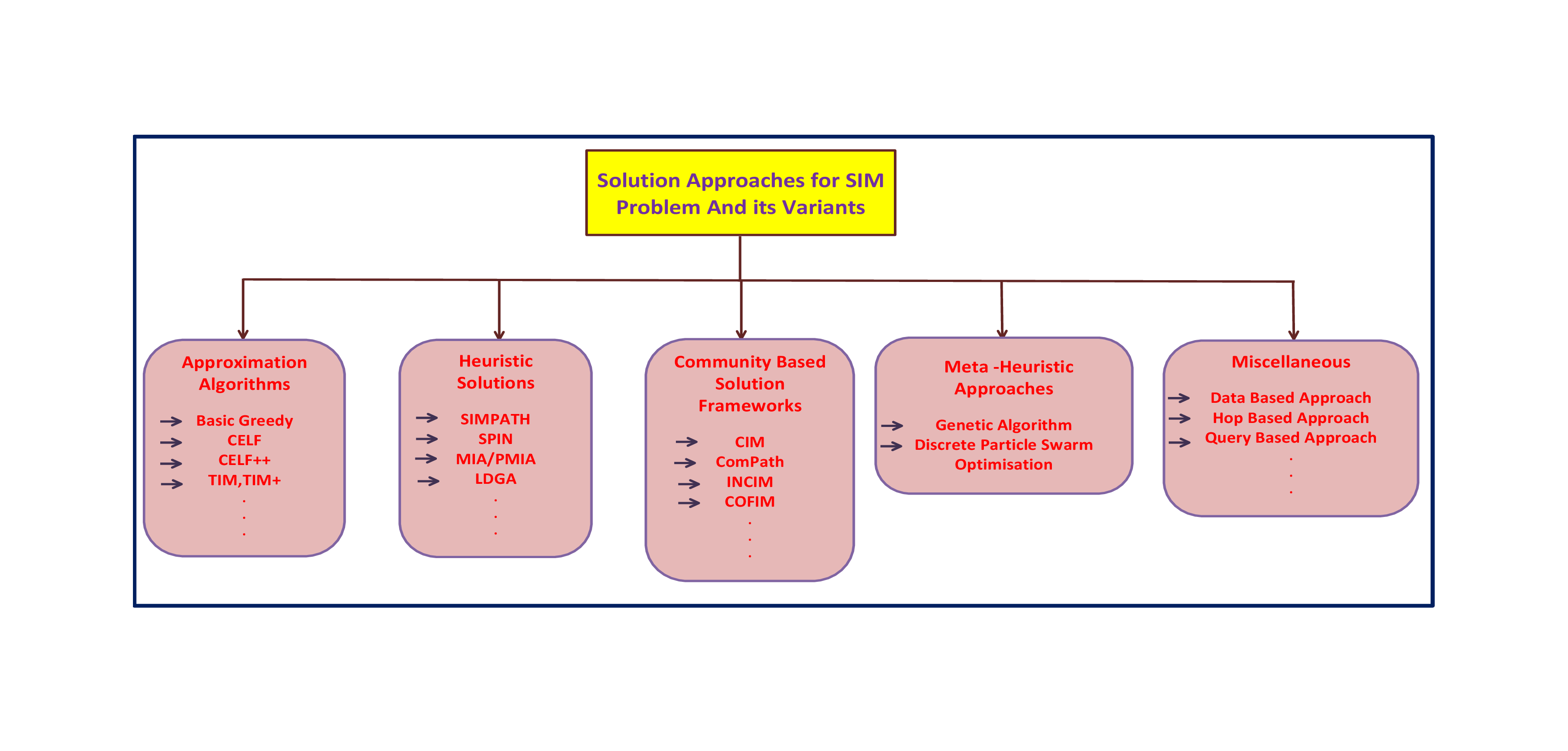}
  \caption{Proposed taxonomy for classifying the solution methodologies.}
  \label{Fig:2}
\end{figure}

 Due to the inherent hardness of the SIM Problem, over the years researchers have developed algorithms for finding seed set for obtaining near\mbox{-}optimal influence spread. In this section, the available solution methodologies in the literature have been described. First we describe our proposed taxonomy for classifying the solution methodologies. Figure \ref{Fig:2} gives a diagrammatic representation of the proposed taxonomy and we describe them below. 
\begin{itemize}
\item  \textbf{Approximation algorithms with provable guarantee}: Algorithms in this category give the worst case bound for influence spread. However, most of them suffer from the scalability issues, which means, with the increase of the network size, running time grows heavily. Many of the algorithms of this category have near optimal asymptotic bounds. 
\item \textbf{Heuristic solutions}: Algorithms of this category do not give any worst case bound on influence spread. However, most of them  have more scalability and better running time compared to the algorithms of previous category.
\item \textbf{Meta\mbox{-}heuristic solutions}: Methodologies of this category are the metaheuristic optimization algorithms and many of them are developed based on the evolutionary computation techniques. These algorithms also do not give any worst case bound on influence spread.
\item \textbf{Community\mbox{-}Based Solutions}: Algorithms of this category use community detection of the underlying social network as an intermediate step to bring down the problem into community level and improves scalability. Most of the algorithms of this category are heuristic and hence, do not provide any worst case bound on influence spread. 
\item \textbf{Miscellaneous}: Algorithms of this category do not follow any particular property and hence, we put them  under this heading.  
\end{itemize}
  
\subsection{Approximation Algorithms with Provable Guarantee} \label{Sec:AAPG}
Kempe et al. \cite{kempe2003maximizing} \cite{kempe2005influential} \cite{kempe2015maximizing} were the first to study the problem of social influence maximization as a \textit{combinatorial optimization} problem and investigated its computational issues under two diffusion models, namely LT and IC models. In there studies, they assumed that the \textit{social influence function}, $\sigma()$ is \textit{sub-modular} and \textit{monotone}. The function $\sigma:2^{V(G)} \rightarrow \mathbb{R}^{+}$ will be sub-modular, if it follows the \textit{diminishing return property}, which means $\forall \  \mathcal{S} \subset \mathcal{T} \subset V(G)$, $u_i \in V(G) \setminus \mathcal{T}$; $\sigma(\mathcal{S} \cup u_i)-\sigma(\mathcal{S}) \geq \sigma(\mathcal{T} \cup u_i)-\sigma(\mathcal{T})$ and $\sigma$ will be monotone, if for any $\mathcal{S} \subset V(G)$ and $\forall u_i \in V(G)\setminus \mathcal{S}$, $\sigma(\mathcal{S} \cup u_i) \geq \sigma(\mathcal{S})$. They proposed a greedy strategy for selecting seed set presented in Algorithm \ref{Brufo}.
\begin{algorithm}[H]
	\KwData{Given Social Network $G(V, E, \theta, \mathcal{P})$ and some $k \in \mathbb{Z}^{+}$.}
	\KwResult{Seed Set for diffusion $\mathcal{S} \subset V(G)$.}
	$\mathcal{S} \leftarrow \phi$\;
	\For{$i=1 $ to $k$}{
		
		$u=\underset{u_i \in V(G)\setminus \mathcal{S}}{argmax} \quad \sigma(\mathcal{S} \cup u_i)-\sigma(\mathcal{S})$\;
		$\mathcal{S} \gets \mathcal{S} \cup u$
	}
	$return \quad \mathcal{S}$
	\caption{Kempe et al.'s \cite{kempe2003maximizing} Greedy Algorithm for \textit{Seed Set Selection}. (\textbf{Basic Greedy})}
	\label{Brufo}
\end{algorithm}
Starting with the empty seed set ($\mathcal{S}$), Algorithm \ref{Brufo} iteratively selects node which is currently not in $\mathcal{S}$, and inclusion of which to $\mathcal{S}$ causes the maximum marginal increment in $\sigma()$. Let us assume that $\mathcal{S}_{i}$ denotes the seed set at $i-th$ iteration of the \textit{`for' loop} in Algorithm \ref{Brufo}. In $(i+1)-th$ iteration, $\mathcal{S}_{i+1}=\mathcal{S}_{i} \cup \{ u \}$, if $\sigma(\mathcal{S} \cup u)-\sigma(\mathcal{S})$ value  becomes the maximum among all $u \in V(G) \setminus \mathcal{S}_{i}$. This iterative process  will be continued until we reach the allowed cardinality of $\mathcal{S}$. Kempe et al. \cite{kempe2003maximizing}  showed that Algorithm \ref{Brufo} provides $(1-\frac{1}{e}-\epsilon)$ with $\epsilon>0$ for the approximation bound on influence spread, maintained in Theorem \ref{Th:10}.
\begin{mythm} \label{Th:10}
	Algorithm \ref{Brufo} provides $(1-\frac{1}{e}-\epsilon)$ with $\epsilon>0$ factor approximation bound for the SIM Problem; i.e.; if $\mathcal{S}^{*}$ be the $k$ element optimal seed set, then $\sigma(\mathcal{S}) \geq (1-\frac{1}{e}).\sigma(\mathcal{S}^{*})$, where $e=\sum_{x=1}^{\infty} \frac{1}{x!}$.
\end{mythm}
Though Algorithm \ref{Brufo} gives good approximation bound on influence spread, it suffers from two major shortcomings. For example, for any given seed set $\mathcal{S}$, exact computation of the influence spread (i.e., $\sigma(\mathcal{S})$) is $\#P\mbox{-}Complete$. Hence, they approximate the influence spread by running a huge number of \textit{Monte Carlo Simulations} (MCS), counting total number of influenced nodes in all simulation runs and taking average with the number of runs. However, recently Maehara et al. \cite{maehara2017exact} developed the first procedure for exact computation of influence spread using \textit{binary decision diagrams}. Secondly, the number of times influence function ($\sigma(.)$) needs to be evaluated is quite huge. For selecting a seed set of size $k$ with $\mathcal{R}$ number of MCS runs in a social network having $n$ nodes and $m$ edges will require $\mathcal{O}(kmn\mathcal{R})$ number of influence function evaluations. Hence, application of this algorithm for a medium size networks (only consisting of 15000 nodes; though real life networks are much larger) appears to be unrealistic \cite{chen2009efficient}, which means that the algorithm is not scalable enough. 
\par In spite of having a few drawbacks, Kempe et al.'s \cite{kempe2003maximizing} study is considered to be the foundational work on the SIM Problem. This study has triggered a vast amount of research in this direction. In most of the cases, the main focus was to reduce the scalability problem incurred by Basic Greedy Algorithm in Kempe et al.'s work. Some of them landed with heuristics, in which the obtained solution could be far away from the optima. Still a few studies are there, in which scalability problem was reduced significantly without loosing approximation ratio. Here, we have listed the algorithms which could provide approximation guarantee, whereas in Section \ref{Sec:HS}, we have described all the  heuristic methods.
\begin{itemize}
\item \textbf{CELF}: For improving the scalability problem, Leskovec et al. \cite{leskovec2007cost} proposed a \textit{Cost Effective Lazy Forward} (CELF) scheme by exploiting the sub-modularity property of the social influence function. The key idea in their study was: for any node, its marginal gain in influence spread in the current iteration cannot be more than its marginal gain in the previous iterations. Using this idea, they were able to make a drastic reduction in the number of evaluations of the influence estimation function ($\sigma(.)$), which leads to significant improvement in running time though the asymptotic complexity remains the same as that of the \textit{Basic Greedy Algorithm} (i.e., $\mathcal{O}(kmn\mathcal{R})$). Reported results in their paper shows that CELF can speed up the computation process upto 700 times compared to Basic Greedy Algorithm on benchmark data sets. This algorithm is also applicable in many other contexts, such as finding informative blogs in a \textit{web blog network}, optimal placement of sensors in a water distribution network for detecting out\mbox{-}breaks etc.
\item \textbf{CELF++}: Goyal et al. \cite{goyal2011celf++} proposed an optimized version of CELF by exploiting the sub-modularity property of social influence function and named it as CELF++. For each node $u$ of the network, CELF++ maintains a table of the form $<u.mg1, u.prev\_ best, u.mg2, u.f\_lag>$ where $u.mg1$ is the marginal gain in $\sigma(.)$ for the current $\mathcal{S}$; $u.prev\_ best$ is the node with the maximum marginal gain among the users scanned till now in the current iteration; $u.mg2$ is the marginal gain in $\sigma(.)$ for $u$ with respect to the $\mathcal{S} \cup \{prev\_ best \}$ and $u.flag$ is the iteration number, when $u.mg1$ was last updated. The key idea in CELF++ is that, if $u.prev\_ best$ is included in the seed set in the current iteration, then the marginal gain of $u$ in $\sigma(.)$ with respect to $\mathcal{S} \cup \{prev\_ best \}$ need not be recomputed in the next iteration. Reported results showed that CELF++ is 35-55 \% faster than CELF though the asymptotic complexity remains the same. 
\item \textbf{Static Greedy}: Cheng et al. \cite{cheng2013staticgreedy} developed this algorithm for solving SIM problem, which provides both guaranteed accuracy as well as high scalability. This algorithm works in two stages. In the first stage, $R$ number of Monte Carlo snapshots are taken from the social network, where each edge $(uv)$ is selected based on the associated diffusion probability $p_{uv}$. In the second stage, starting from the empty seed set, a node having the maximum average marginal gain in influence spread over all sampled snapshots will be selected as a seed node. This process will be continued until $k$ nodes are selected. This algorithm has the running time of $\mathcal{O}(\mathcal{R}m + k\mathcal{R}m^{'}n)$ and space requirement of $\mathcal{O}(\mathcal{R}m^{'})$, where $\mathcal{R}$ and  $m^{'}$ are the number of Monte Carlo samples and average number of active edges in the snapshots, respectively. Reported results show that the Static Greedy reduces the computational time by two orders of magnitude, while achieving the better influence spread compared to Degree Discount Heuristic (DDH), Maximum Degree Heuristic (MDH), Prefix excluding Maximum Influence Arborescence (PMIA) (discussed in Section \ref{Sec:HS}) Algorithms.
\item \textbf{Borgs et al.'s Method:} Borgs et al. \cite{borgs2014maximizing} proposed a completely different approach for solving SIM Problem under IC Model using \textit{reverse reachable sampling technique}. Other than the MCS runs , this is a new approach for estimating the influence spread. Their algorithm is randomized and succeeds with the probability of $\frac{3}{5}$ and has the running time of $\mathcal{O}((m+n)\epsilon^{-3}\log n)$, which  improves the previously best known algorithm having the complexity of $\mathcal{O}(mnkPOLY(\epsilon^{-1}))$. Algorithm proposed by Borgs et al. is near\mbox{-}optimal since the lower bound is $\Omega(m+n)$. This algorithm works in two phases. In the first phase, stochastically a hypergraph ($\mathcal{H}$) is generated from the input social network. Second phase is concerned with the seed set selection. This is done by repeatedly choosing the node with maximum degree in $\mathcal{H}$, deleting it along with its incidence edges from $\mathcal{H}$. The $k$\mbox{-}element set obtained in this way is the seed set for diffusion. This work is mostly theoretically enriched and lacking of practical experimentation.
\item \textbf{Zohu et al.'s Method}: Zohu et al. \cite{zhu2015better} improved the approximation bound from $(1-\frac{1}{e})$ (which is approximately 0.63) to 0.857. They designed two approximation algorithms: first algorithm works for the problem, where the cardinality of the seed set ($\mathcal{S}$) is not restricted and the second one works, when there is some restricted upper bound on the cardinality of seed set. They formulated the influence maximization problem as an optimization problem given below.
\begin{equation}
\underset{\mathcal{S} \subset V(G)}{max} \quad \underset{u \in \mathcal{S}, v \in V(G) \setminus \mathcal{S}}{\sum} p_{uv},
\end{equation}
where $p_{uv}$ is the \textit{influence probability} between the users:  $u$ and $v$. They converted this optimization problem into a \textit{quadratic integer programming problem} and solved the problem using the concept of \textit{semidefinite programming} \cite{feige1995approximating}. 
\item \textbf{SKIM}: Cohen et al. \cite{cohen2014sketch} proposed a \textit{Sketch\mbox{-}Based Influence Maximization} (SKIM) algorithm, which improves the Basic Greedy Algorithm by ensuring in every iteration, with sufficiently high probability, or in expectation, the node we choose to add to the seed set has a marginal gain that is close to the maximum one. The running time of this algorithm is $\mathcal{O}(nl+ \sum_{i=1} \vert E^{i} \vert + m \epsilon^{-2} \log^{2}n)$, where $l$ is the number of snap shots of $G$, $E^{i}$ is the edge set of $G^{i}$. Reported results show that SKIM has high scalability over Basic Greedy, Two phase Influence Maximization (TIM), Influence Ranking and Influence Estimation (IRIE) etc. without compromising influence spread.  
\item \textbf{TIM}: Tang et al. \cite{tang2014influence} developed a \textit{Two\mbox{-}phase Influence Maximization} (TIM) algorithm, which has the expected running time of $\mathcal{O}((k+l)(n+m)\log n/ \epsilon^2)$ with atleast $(1-n^{-l})$ probability for some given $k$, $\epsilon$ and $l$. As its name suggests, this algorithm has two phases. In the first phase, TIM computes lower bound on the maximum expected influence spread among all $k$ sized sets and uses this lower bound to estimate a parameter $\phi$. In the second phase, $\phi$ number of reverse reachability (RR) set samples have been picked up from the social network. Then, it derives a $k$ sized seed set that covers the maximum number of RR sets and returns as the final result. Reported results shows that TIM is two times faster than CELF++ and Borgs et al.'s \cite{borgs2014maximizing} Method, while achieving the same influence spread. To improve the running time of TIM, Tang et al. \cite{tang2014influence} proposed a heuristic, which takes all the RR sets, generated in an intermediate step of second phase of TIM as inputs. Then, it uses a greedy approach for the maximum coverage problem for selecting the seed set. This modified version of TIM is named as $\text{TIM}^{+}$. Reported results showed that $\text{TIM}^{+}$ is two times faster than TIM.
\item \textbf{IMM}: Tang et al. \cite{tang2015influence} proposed \textit{Influence Maximization via Martingales} (IMM) (a kind of stochastic process, in which, for the given current and preceding values, the conditional expectation of the next value, will be the current value itself), which achieves a $\mathcal{O}((k+l)(n+m)\log n/ \epsilon^2)$ expected running time and returns $(1-\frac{1}{e} - \epsilon)$ factor approximate solution with probability of $(1-n^{-l})$. IMM Algorithm also has two phases like TIM and $\text{TIM}^{+}$. First phase is concerned with sampling $RR$ sets from the given social network and the second phase is concerned with the seed set selection. In the first phase, unlike TIM and $\text{TIM}^{+}$, RR sets generated in the first phase are dependent because $(i+1)$\mbox{-}th RR set is generated based on whether first $i$ of RR sets are satisfying stopping criteria or not. In IMM, the RR sets generated in the sampling phase are reused in node selection phase, which is not the case in TIM or TIM+. In this way, IMM can eliminate a lot of unnecessary computations, which leads to significant improvement in running time though asymptotic complexity remains the same as that of TIM. Reported results conclude that IMM outperforms TIM, TIM+, IRIE (described in Section \ref{Sec:HS}) based on running time while achieving comparable influence spread.
\item \textbf{Stop-and-Stare}: Nguyen et al. \cite{nguyen2016stop} developed the Stop-and-Stare Algorithm (SSA) and its dynamic version DSSA for \textit{Topic\mbox{-}aware Viral Marketing} (TVM) problem. We have not discussed this problem, as it comes under topic aware influence maximization. However, this solution methodology can be used for solving SIM problem with minor modification. They showed that, the number of RR set samples used by their algorithms is asymptotically minimum. Hence, Stop-and-Stare is 1200 times faster than the state-of-the art IMM algorithm. We are not discussing the results, as they are for the TVM problem and out of the scope of this survey.  
\item \textbf{BCT}: Recently, Nguyen et al. \cite{nguyen2017billion} proposed \textit{Billion-scale Cost-award Targeted} (BCT) algorithm for solving \textit{cost-aware targeted viral marketing} (CTVM) introduced by them. We have not discussed this problem, as it comes under topic aware influence maximization. However, this solution methodology can be adopted for solving SIM Problem as well under both IC and LT Models and have the running time of $\mathcal{O}((k+l)(n+m)\log n/ \epsilon^2)$ and $\mathcal{O}((k+l)n\log n/ \epsilon^2)$, respectively. We are not discussing about the results, as they are for CTVM Problem and out of scope of this survey.
  
\item \textbf{Nguyen et al.'s Method}: Nguyen et al. \cite{nguyen2013budgeted} studied the \textit{Budgeted Influence Maximization Problem} described in Section \ref{Sec:VTSSP}. They have formulated the following optimization problem in the context of \textit{Budgeted Influence Maximization}:\\
\begin{eqnarray}
max \quad \sigma(\mathcal{S}) \\
\text{subject to,}\underset{u \in \mathcal{S}}{\sum} \mathcal{C}(u) \leq \mathcal{B}
\end{eqnarray}
Now, if $\forall u \in V(G)$, $\mathcal{C}(u)=1$, then it becomes the SIM Problem. To solve this problem, they proposed two algorithms. First one is the modification of basic greedy algorithm proposed by Kempe et al. \cite{kempe2003maximizing} (Algorithm \ref{Brufo}) and second one was adopted from \cite{khuller1999budgeted}. In the first algorithm  $\forall u \in V(G)\setminus \mathcal{S}$, they computed the increment of influence in unit cost as follows:
\begin{equation}
\delta(u)=\frac{\sigma(\mathcal{S} \cup u)- \sigma(\mathcal{S})} {\mathcal{C}(u)}
\end{equation}
Now, the algorithm choose $u$ to include in the seed set ($\mathcal{S}$), if it maximized the \textit{objective function} as well as $\mathcal{C}(\mathcal{S}_{i} \cup u) \leq \mathcal{B}$. This iterative process will be continued until no more nodes can be added within the budget. However, this algorithm does not give any constant approximation ratio. This algorithm can be modified to get the constant approximation ratio, as given in Algorithm \ref{Algo:2}.

\begin{algorithm}[H]
	\KwData{Given Social Network $G(V, E, \theta, \mathcal{P})$, cost function $\mathcal{C}: V(G) \longrightarrow \mathbb{Z}^{+}$ some $\mathcal{B} \in \mathbb{Z}^{+}$.}
	\KwResult{Seed Set for diffusion $\mathcal{S} \subset V(G)$.}
		$S_{1}= \text{result of Naive Greedy}$\;
		$S_{max}= \underset{u \in V(G)}{argmax} \quad \sigma(u)$\;
	    $\mathcal{S}=argmax(\sigma(S_{1}), \sigma(S_{max}))$\;
	    $return \quad \mathcal{S}$
	\caption{Nguyen et al.'s \cite{nguyen2013budgeted} Greedy Algorithm for BIM Problem.}
	\label{Algo:2}
\end{algorithm}
\begin{mythm}
  Algorithm \ref{Algo:2} guarantees $(1-\frac{1}{\sqrt{e}})$ approximate solution for \textit{BIM Problem}.
\end{mythm}
For the detailed proof of Algorithm \ref{Algo:2}, readers are referred to the appendix of \cite{nguyen2012budgeted}. 
\end{itemize}
\par Now, the presented algorithms have been summarized below. The main bottleneck in Kempe et al.'s \cite{kempe2003maximizing} Basic Greedy Algorithm is the evaluation of influence spread estimation function for a large number of MCS runs (say, 10000). If we reduce the MCS runs directly, then accuracy in computing influence spread may be compromised. So, the key scope for improvement is to reduce the number of evaluation of the influence estimation function in each MCS run. Both CELF and CELF++ exploit the sub-modularity property to achieve this goal and hence, are found to be faster than Basic Greedy Algorithm. On the other hand, Static Greedy algorithm uses all the randomly generated snapshots of the social network using MCS runs simultaneously. Hence, with the less number of MCS runs (say, 100) it is possible to have equivalent accuracy in spread. These four algorithms can be ordered in terms of maximum to minimum values of running time as follows: $\text{Basic Greedy} \succ \text{CELF} \succ \text{CELF++} \succ \text{Static Greedy}$.
\par Another scope of improvement in Kempe et al.'s \cite{kempe2003maximizing} work was  estimating the influence spread by applying some method other than the heavily time consuming MCS runs. Borgs et al. \cite{borgs2014maximizing} explored this scope by proposing a drastically different approach for spread estimation, namely reverse reachable sampling technique. The algorithms (such as TIM, $\text{TIM}^{+}$, IMM) which used this method were seem to be much faster than CELF++ and also have competitive influence spread. Among TIM, $\text{TIM}^{+}$, and IMM , IMM was found to be the fastest one both theoretically (in terms of computational complexity), and empirically (in terms of computational time from experimentation) due to the reuse of the RR sets in the node selection phase. To the best of  the authors' knowledge, IMM is the fastest algorithm, which was solely proposed for solving SIM Problem. However, BCT Algorithm proposed by Nguyen et al. \cite{nguyen2017billion}, which was originally proposed for solving CTVM problem, is the fastest solution methodology available in the literature that can be adopted for solving SIM Problem.
\par Now from this discussion, it is important to note that the scalability problem incurred by the Basic Greedy Algorithm had been reduced by the subsequent research. However, as the size of the social network data set has become gigantic, development of algorithms with high scalability remains the thrust area. Solution methodologies described till now have been summarized in Table \ref{Tab:4}. Algorithms for which complexity analysis had not been done by the author(s), we left that column of the table blank.
\begin{center}
\begin{table} \label{Tab:4}
    \begin{tabular}{ | p{2.2 cm} | p{1.8 cm} | p{2.9cm} | p{2cm}| p{1 cm} |}
    \hline
    \textbf{Name of the Algorithm} & \textbf{Proposed By} & \textbf{Complexity} & \textbf{Applicable For} & \textbf{Model} \\ \hline
    \textbf{Basic Greedy} & Kempe et al. \cite{kempe2003maximizing} & $\mathcal{O}(kmn\mathcal{R})$  &  \textbf{SIM} & IC \& LT\\
    \hline
    \textbf{CELF} & Leskovec et al. \cite{leskovec2005graphs} & $\mathcal{O}(kmn\mathcal{R})$  & \textbf{SIM} & IC \& LT\\
    \hline
    \textbf{CELF++} & Goyal et al.\cite{goyal2011celf++} & $\mathcal{O}(kmn\mathcal{R})$  & \textbf{SIM} & IC \& LT \\ 
    \hline
    \textbf{Static Greedy} & Cheng et al. \cite{cheng2013staticgreedy} & $\mathcal{O}(\mathcal{R}m + kn\mathcal{R}m)$ & \textbf{SIM} & IC \& LT \\
    \hline
    \textbf{Brog et al.'s Method} & Brogs et al. \cite{borgs2014maximizing} & 
    $\mathcal{O}(kl^2(m+n)\log^2n/\epsilon^3)$ & \textbf{SIM} & IC \& LT \\
    \hline
    \textbf{Zohu et al.'s Method} & Zohu et al. \cite{zhu2015better} & - & \textbf{SIM} & IC \& LT \\
    \hline
    \textbf{SKIM} & Cohen et al. \cite{cohen2014sketch} & $\mathcal{O}(nl+ \sum_{i=1} \vert E^{i} \vert + m \epsilon^{-2} \log^{2}n)$ & \textbf{SIM} & IC \& LT \\
    \hline
    \textbf{TIM+}, \textbf{IMM} & Tang et al. \cite{tang2014influence}, \cite{tang2015influence} & $\mathcal{O}((k+l)(n+m)\log n/ \epsilon^2)$  & \textbf{SIM} & IC \& LT \\
    \hline
    \textbf{Stop-and-Stare} & Nguyen et al. \cite{nguyen2016stop} & - & \textbf{TVM} & IC \& LT \\ 
    \hline
   \textbf{Nguyen's Method} & Nguyen et al. \cite{nguyen2013budgeted} &  $\mathcal{O}(n^2 (\log n+d) + kn(1+d))$ &\textbf{BIM} &  IC \& LT \\
    \hline
    \textbf{BCT} & Nguyen et al. \cite{nguyen2017billion} & $\mathcal{O}((k+l)(n+m)\log n/ \epsilon^2)$& \textbf{SIM}, \textbf{BIM}, \textbf{CTVM} & IC \\
    \hline
    \textbf{BCT} & Nguyen et al. \cite{nguyen2017billion} & $\mathcal{O}((k+l)n\log n/ \epsilon^2)$& \textbf{SIM}, \textbf{BIM}, \textbf{CTVM} & LT \\
    \hline
    \end{tabular}
    \caption{Approximation algorithms for SIM Problem and its variants.}
\end{table}
\end{center}


\subsection{Heuristic Solutions} \label{Sec:HS}
 Algorithms of this category do not provide any approximation bound on the influence spread but have better running time and scalability. Here, we will describe the heuristic solution methodologies from the literature.
\begin{itemize}
    \item \textbf{Random Heuristic}: For selecting seed set by this method, randomly pick $k$ nodes of the network and return them as seed set. In Kempe et al.'s \cite{kempe2003maximizing} experiment, this method has been used as a baseline method. 
	\item \textbf{Centrality\mbox{-}Based Heuristics}: Centrality is a well\mbox{-}known measure in network analysis, which signifies how much importance a node has in the network \cite{freeman1978centrality} \cite{landherr2010critical}. There are many centrality\mbox{-}based heuristics proposed in the literature for SIM Problem like \textit{Maximum Degree Heuristic} (MDH) (select $k$ highest degree nodes as seed node), High Clustering Coefficient Heuristic (HCH) (select $k$ nodes with the highest clustering coefficient value) \cite{wilson2009user} \cite{tabak2014directed}, High page rank heuristic \cite{Brin98The} (select $k$ nodes with the highest page rank value) etc.
	\item \textbf{Degree Discount Heuristic} (DDH): This is basically the modified version of MDH and was proposed by Chen et al. \cite{chen2009efficient}. The key idea behind this method is following for any two nodes $u,v \in V(G)$, $(uv) \in E(G)$ and $u$ has been selected as a seed set by MDH, and then, during the counting the degree of $v$, the edge $(uv)$ should not be considered. Hence, due to the presence of $u$ in the seed set, the degree of $v$ will be discounted by 1. This method is also named as \textit{Single Discount Heuristic} (SDH).  Experimental results of \cite{chen2009efficient} show that DDH can achieve better influence spread than MDH. 
	\item \textbf{SIMPATH}: This heuristic was proposed by Goyal et al. \cite{goyal2011simpath} for solving SIM Problem under LT Model. SIMPATH works based on the principal of CELF (discussed in Section \ref{Sec:AAPG}). However, instead of using computationally expensive Monte Carlo Simulations for estimating influence spread, SIMPATH uses path enumeration techniques for this purpose. This algorithm has a parameter ($\eta$) for controlling trade off between influence spread and running time. Reported results conclude that SIMPATH outperforms other heuristics, such as MDH, Page Rank, LDGA with respect to information spread. 
	\item \textbf{SPIN}: Narayanam et al. \cite{narayanam2011shapley} studied SIM Problem and $\lambda$ Coverage Problem as a co\mbox{-}operative game and proposed a \textit{Shapely Value\mbox{-}Based Discovery of Influential Nodes} (SPIN) Algorithm, which has the running time of $\mathcal{O}(t(n+m)\mathcal{R} + n \log n+ kn+ k\mathcal{R}m)$, where $t$ is the cardinality of the sample collision set being considered for the computation of shapely value. This algorithm has mainly two steps. First one is to generate a rank list of the nodes based on the shapley value and then, choose top\mbox{-}k of them and return as seed set. Reported results show that SPIN constantly outperforms MDH and HCH.
	\item \textbf{MIA} and \textbf{PMIA}: Chen et al. \cite{chen2010scalable} and Wang et al. \cite{wang2012scalable} proposed \textit{maximum influence arborescence} (MIA) and Prefix excluding MIA (PMIA) model of influence propagation. They computed the propagation probability from a seed node to a non\mbox{-}seed node by multiplying the influence probabilities of the edges present in the shortest path. \textit{Maximum Influence Path} is the one having the maximum propagation probability and they considered that influence spreads through local arborescence (a directed graph in which, for a vertex $u$ called the root and any other vertex $v$, there is exactly one directed path from $u$ to $v$) only. Hence, the model is called MIA. In PMIA (\textit{Prefix excluding} MIA) model, for any seed $s_i$, its maximum influence path to other nodes should avoid all seeds that are before $s_i$. They proposed greedy algorithms for selecting seed set based on these two diffusion models. Reported results show that both MIA and PMIA can achieve high level of scalability. 
	\item \textbf{LDAG}: Chen et al. \cite{chen2010scalable2} developed  this heuristic for solving SIM Problem under LT Model. Influence spread in a  \textit{Directed Acyclic Graph} (DAG) is easy to compute. Hence, for computing the influence spread in general social networks, they  introduced a \textit{Local Directed Acyclic Graph} (LDAG) based influence model, which computes local DAGs for each node to approximate influence spread. After constructing the DAGs, basic greedy algorithm proposed by Kempe et al. \cite{kempe2003maximizing} can be used to select the seed nodes. Reported results show that LDAG constantly outperforms DDH or Page Rank heuristic. 
\item \textbf{IRIE}: Jung et al. \cite{jung2012irie} proposed this heuristic based on influence ranking (IR) and influence estimation (IE) for solving SIM Problem under IC and its extension IC\mbox{-}N (independent cascade with negative opinion) Model. They developed a global influence ranking like belief propagation approach. If we select top\mbox{-}k nodes, then there will be an overlap in influence spread by each node. For avoiding this shortcomings, they integrated a simple \textit{influence estimation} technique to predict additional influence impact of a seed on the other node of the network. Reported results show that IRIE can achieve better influence spread compared to MDH, Pagerank, PMIA etc. heuristics. However, IRIE has less running time and memory consumption.
\item \textbf{ASIM}: Galhotra et al. \cite{galhotra2015asim} designed  this highly scalable heuristic for SIM Problem. For each node $u \in V(G)$, this algorithm assigns a score value (the weighted sum of the number of simple paths of length at most $d$ starting from that node). ASIM has the running time of $\mathcal{O}(kd(m+n))$ and its idea is quite similar to the SIMPATH Algorithm proposed by Goyal et al. \cite{goyal2011simpath}. Results show that ASIM takes less computational time and consumes less memory compared to CELF++ and TIM, while achieving the comparable influence spread.  
\item \textbf{EaSyIm}: Galhotra et al. \cite{galhotra2016holistic} proposed \textit{opinion cum interaction} (OCI) model, which considers negative opinion as well. Based on the OCI Model, they formulated the \textit{maximizing effective opinion} problem and proposed two fast and scalable heuristics, namely Openion Spread Influence Maximization (OSIM) and EaSyIm having the running time of $\mathcal{O}(k \mathcal{D}(m+n))$ for this problem, where $\mathcal{D}$ is the diameter of the graph. Both the algorithms work in two phases. In the first phase, each node is assigned with some score based on the contribution on influence spread for all the paths starting at that node. Second step is concerned with the node processing step. The nodes with the maximum score value are selected as seed nodes. Reported empirical results show that OSIM and EaSyIm can achieve better influence spread compared to $\text{TIM}^{+}$, CELF++ with less running time.
	\item \textbf{Cordasco et al.'s \cite{cordasco2015fast} \cite{cordasco2016active} Method}: Later Cordasco et al. proposed a fast and effective heuristic method for selecting the target set in a undirected social network \cite{cordasco2015fast} \cite{cordasco2016active}. This heuristic produces optimal solution for \textit{trees}, \textit{cycles} and \textit{complete graphs}. However, for real life social networks, this heuristic performs much better than the other methods available in the literature. They extended this work for directed social networks as well \cite{cordasco2015influence}. 
\end{itemize}	
 There are several other studies also, which focused on developing heuristic. Nguyen et al. \citep{nguyen2013budgeted} proposed an efficient heuristic for solving BIM Problem.  Wu et al. \cite{wu2017two} developed a two\mbox{-}stage stochastic programming approach for solving SIM Problem. In this study, instead of choosing a seed set of size exactly $k$, their problem is choosing a seed set of size less than or equal to $k$.
\par Now, the studies related to heuristic methods will be summarized here. Centrality\mbox{-}based heuristics (CBHs) consider the topology of the network only and hence, obtained influence spread in most of the cases is quite less compared to that of other states of the art methods. However, DDH performs slightly better than other CBHs, as it puts a little restriction on the selection of two adjacent nodes. The application of SIMPATH for seed selection is little advantageous, as it has a user controlled parameter $\eta$ to balance the trade\mbox{-}off between accuracy and running time. SPIN has the advantage, as it can be used for solving both Top\mbox{-}$k$ node problem as well as $\lambda$\mbox{-}Coverage Problem. MIA and PMIA have the better scalability compared to Basic Greedy. As LDAG works based on the principle of computation of influence spread in DAGs, it is seen to be faster. As various heuristics are experimented with different benchmark data sets, drawing a general conclusion about the performance will be difficult. Here, we have summarized some of the important algorithms for solving SIM and related problems, as presented in Table \ref{Tab:HS}. Algorithms for which complexity analysis has not been done in the paper, we have left that column empty in the table.
\begin{table} [h]
\begin{center}
    \begin{tabular}{ | p{2cm} | p{2.5 cm} | p{2.7cm} | p{1 cm} |}
    \hline
    \textbf{Name of the Algorithm} & \textbf{Proposed By} & \textbf{Complexity}  & \textbf{Model} \\ \hline
    \textbf{SIMPATH} & Goyal et al. \cite{goyal2011simpath} & $\mathcal{O}(kmn\mathcal{R})$ & LT \\
    \hline
    \textbf{SPIN} & Narayanam et al. \cite{narayanam2011shapley} & $\mathcal{O}(t(n+m)\mathcal{R} + n \log n+ kn+ k\mathcal{R}m)$ & IC \& LT \\
    \hline
    \textbf{MIA},\textbf{PMIA} & Chen et al. \cite{chen2010scalable}, Wang et al. \cite{wang2012scalable} & - & MIA, PMIA\\
    \hline
    \textbf{LDGA} & Chen et al. \cite{chen2010scalable} & $\mathcal{O}(n^2 + k n^2 \log n)$ & MIA \\
    \hline
    \textbf{IRIE} & Jung et al. \cite{jung2012irie} & - & IC \& IC\mbox{-}N \\
    \hline
    \textbf{ASIM} & Galhotra et al. \cite{galhotra2015asim} & $\mathcal{O}(kd(m+n))$ & IC \\
    \hline
    \textbf{EaSyIm} & Galhotra et al. \cite{galhotra2016holistic} & $\mathcal{O}(k\mathcal{D}(m+n))$ & OI \\
    \hline
    \end{tabular}
\end{center}
\caption{Heuristic solutions for SIM Problem}
\label{Tab:HS}
\end{table}
\subsection{Metahuristic Solution Approaches} \label{Sec:Meta}
Since early seventies, metaheuristic algorithms had been used successfully to solve optimization problems arises in the broad domain of science and engineering \cite{yi2013three} \cite{yang2014computational}. There is no exception for solving SIM Problem as well.
\begin{itemize} 
\item Bucur et al. \cite{bucur2016influence} solved the SIM Problem using \textit{genetic algorithm}. They demonstrated that with simple genetic operator, it is possible to find out approximate solution for influence spread within feasible run time. In most of the cases, influence spread obtained by their method was comparable with that of the Basic Greedy Algorithm proposed by Kempe et al. \cite{kempe2003maximizing}.
 \item Jiang et al. \cite{jiang2011simulated} proposed \textit{simulated annealing}\mbox{-}based algorithm for solving the SIM Problem under IC Model. Reported results indicate that their proposed methodology runs 2-3 times faster compared to the existing heuristic methods in the literature. 
 \item Tsai et al. \cite{tsai2015genetic} developed the \textit{Genetic New Greedy Algorithm} (\textbf{GNA}) for solving SIM Problem under IC Model by combining genetic algorithm with the new greedy algorithm proposed by Chen et al. \cite{chen2009efficient}. Their reported results conclude that GNA can give 10 \% more influence spread compared to the genetic algorithm. 
\item Gong et al. \cite{gong2016influence} proposed a \textit{discrete particle swarm optimization algorithm} for solving SIM Problem. They used the degree discount heuristic proposed by Chen et al. \cite{chen2009efficient} to initialize the seed set and \textit{local influence estimation (LIE) function} to approximate the two-hop influence. They introduced the \textit{network specific local search} strategy also for fast convergence of their proposed algorithm. Reported results conclude that this methodology outperforms the state of the art CELF++ with less computational time. 
\end{itemize}
After that, several studies were also carried out in this direction \cite{sankar2016learning}, \cite{wang2017discrete}, \cite{liu2017effective} \cite{zhang2017maximizing}. Though there are a large number of metaheuristic algorithms  \cite{yang2010nature}, only a few had been used for solving SIM Problem. Hence, the use of metaheuristic algorithms for solving SIM Problem and its variants has been largely ignored. Next, we have described the community\mbox{-}based solution methodologies for SIM Problem. 
\subsection{Community\mbox{-}Based Solution Approaches} Most of the real\mbox{-}life social networks exhibit a community structure within it \cite{clauset2004finding}. A community is basically a subset of nodes, which are densely connected among themselves and sparsely connected with the other nodes of the network. In recent years, \textit{community\mbox{-}based solution framework} (\textbf{CBSF}) has been developed for solving SIM Problem. 
\begin{itemize}
\item Wang et al. \cite{wang2010community} proposed the \textit{community\mbox{-}based greedy algorithm} for solving SIM Problem. This method consist of two steps, namely detecting communities based on information propagation and     selecting communities for finding influential nodes. This algorithm could outperform the degree discount and random heuristic. 
\item Chen et al. \cite{chen2012exploring} \cite{chen2014cim} developed a CBSF for solving SIM Problem and named it \textbf{CIM}. By exploiting the community structure, they selected some candidate seed sets, for each community and from the candidate seed sets they have selected the final seed set for diffusion. CIM could achieve better influence spread compared to some state\mbox{-}of\mbox{-}the art heuristic methods, such as CDH-Kcut, CDH-SHRINK and maximum degree. \item Rahimkhan et al.  \cite{rahimkhani2015fast} proposed a CBSF for solving SIM Problem under LT Model and named it \textbf{ComPath}. They used Speaker-
listener Label Propagation Algorithm (SLPA) proposed by  Xie et al. \cite{xie2011slpa} for detecting communities and then identified the most influential communities and candidate seed nodes. From the candidate seed set, they selected the final seed set based on the intra distance among the nodes of the candidate seed set. ComPath could outperform CELF, CELF++, maximum degree heuristic, maximum pagerank heuristic, LDGA. 
\item Bozorgi et al. \cite{bozorgi2016incim} developed a CBSF for solving SIM Problem under LT Model and named it \textbf{INCIM}. Like ComPath, INCIM also use the SLPA Algorithm for detecting the communities. They proposed an algorithm for selecting seed, which computes the influence spread using the algorithm developed by Goyal et al. \cite{goyal2011simpath}. INCIM could outperform some state-of-the-art methodologies like LDGA, SIMPATH, IPA (a parallel algorithm for SIM Problem proposed by \cite{kim2013scalable}),  high pagerank and high degree heuristic. 
\item Shang et al. \cite{shang2017cofim} proposed a CBSF for solving SIM Problem and named it \textbf{CoFIM}. In this study they introduced a diffusion model, which works in two phases. In the first phase the seed set $\mathcal{S}$ was expanded to the neighbor nodes of $\mathcal{S}$, which would be usually allocated into different communities. Then, in the second phase, influence propagation within the communities was computed. Based on this diffusion model, they developed an incremental greedy algorithm for selecting seed set, which is analogous to the algorithm proposed by Kempe et al. \cite{kempe2003maximizing}. CoFIM could achieve better influence spread compared to that of IPA, TIM+, MDH and IMM.
\item Recently, Li et al. \cite{li2018community} proposed a community\mbox{-}based approach for solving the SIM Problem, where the users have a specific geographical location. They developed a social influence\mbox{-}based community detection algorithm using spectral clustering technique and a seed selection methodology by considering community\mbox{-}based influence index. Reported results show that this methodology is more efficient than many state\mbox{-}of\mbox{-}the\mbox{-}art methodologies, while achieving almost the same influence spread.
\end{itemize}
\par It is important to note that except the methodology proposed by Wang et al. \cite{wang2010community}, all these methods are  basically heuristics. However, these methods use community detection of the underlying social network as an intermediate step to scale down the SIM Problem into community level. There are large number of algorithms available in the literature for detecting communities \cite{fortunato2010community}, \cite{chakraborty2017metrics}. Among them, which one should be used for solving SIM Problem? How is the quality of community detection and influence spread  related? This questions are largely ignored in the literature.   
\subsection{Miscellaneous} In this section, we have described some solution methodologies of SIM Problem, which are very different from the methodologies discussed till now. Also, each solution methodology presented here is different from another. It is reported in the literature that in any information diffusion process less than 10\% nodes are influenced beyond the hop count $2$ \cite{goel2012structure}. Based on this phenomenon, recently, Tang et al. \cite{tang2017influence} \cite{tang2018efficient} developed  a hop\mbox{-}based approach for SIM Problem. Their methodology also gives a theoretical guarantee on influence spread. Ma et al. \cite{ma2008mining} proposed an algorithm for SIM Problem, which works based on the heat diffusion process. It could produce better influence spread compared to Basic Greedy Algorithm. Goyal et al. \cite{goyal2011data} developed a data\mbox{-}based approach for solving SIM Problem. They introduced the \textit{credit distribution (CD) model} that could grip the propagation traces to learn the influence flow pattern for approximating the influence spread. They showed that SIM Problem under CD Model is NP\mbox{-}Hard and reported results show that this model can achieve even better influence spread compared to IC and LT Models with less running time. Lee et al. \cite{lee2015query} introduced a query\mbox{-}based approach for solving SIM Problem under IC Model. Here, the query is for activating all the users of a given set $\mathcal{T}$, what should be the seed set?  This methodology is intended for maximizing the influence of a particular group of users, which is the case in \textit{target-aware viral marketing}. Zhu et al. \cite{zhu2014maximizing} introduced the \textbf{CTMC\mbox{-}ICM} diffusion model, which is basically the blending of IC Model with \textit{Continuous Time Markov Chain}. They studied the SIM Problem under this model and came up with a new centrality metric  \textit{Spread Rank}. Their reported results show that seed nodes selected based on spread rank centrality can achieve better influence spared compared to the traditional distance\mbox{-}based centrality measures, such as \textit{degree}, \textit{closeness}, \textit{betweenness}. Wang et al. \cite{wang2017maximizing} proposed the methodology \textbf{Fluidspread}, which works based on fluid dynamic principle and can reveal the dynamics of diffusion process. Kang et al. \cite{kang2016diffusion} introduced the notion of diffusion centrality for selecting influential nodes in a social network. 

\section{Summary of the Survey and Future Research Directions} \label{Sec:SRD}
Based on the survey of the existing literature presented in Sections \ref{Sec:VTSSP} through \ref{Sec:SolTSS}  we have summarized in this section the current research trends and given future directions.
\subsection{Current Research Trends}
\begin{itemize}
\item \textbf{Practicality of the Problem}: Most of the current studies is focused on the practical issues of the SIM Problem. One of the major applications of social influence maximization is viral marketing. So, in this context, influencing an user will be beneficial, only if he will be able to influence a reasonable number of other users of the network. Recent studies, such as \cite{nguyen2016cost} \cite{nguyen2017billion} along with the node selection cost also consider \textit{benefit} as another component in the SIM problem.
\item \textbf{Scalability}: Starting from kempe et al.'s \cite{kempe2003maximizing} seminal work, scalability remains an important issue in this area. To reduce scalability problem, instead of using Monte Carlo simulation\mbox{-}based spread estimation, recently  Borgs et al. \cite{borgs2014maximizing} introduced reverse reachable set\mbox{-}based spread estimation. After this work, all the popular algorithms for SIM Problem, such as TIM, IMM, TIM+ etc uses this concept as an influence spread estimation technique for improving scalability.
\item \textbf{Diffusion Probability Computation}: TSS problem assumes that influence probability between any pair of users is known. However, this is a very unrealistic assumption. Though there were some previous studies in this direction, people tried to predict influence probability using machine learning techniques \cite{varshney2017predicting}.           
\end{itemize}
Though since the last one and half decades or so, the \textit{TSS Problem} had been studied extensively from both theoretical as well as applied context, still to the best of our knowledge, some of the corners of this problem are either not or partially investigated. Here, we have listed some future research directions from both problem specification as well as solution methodology point of view.
\subsection{Future Directions} 
Further research may be carried out in future in and around of TSS Problem of social networks, in the following directions:
\subsubsection{Problem Specific}
\begin{itemize}
\item As on\mbox{-}line social networks are formed by the rational agents, incentivization is required, if a node is selected as a seed node. For practical applications, it is also important to consider what benefit will be  obtained (e.g., how many other non\mbox{-}seed nodes becoming influenced through that node etc.) by activating that node. At the same time , for influence propagation of time sensitive events ( where influencing one person after an event does not make any scene such as, political campaign before election, viral marketing for a seasonal product etc.) consideration of diffusion time is also important. To the best of our knowledge, there is no reported study on TSS Problem considering all three issues: \textit{cost, benefit, and time}.
\item Most of the studies done on SIM Problem and its variants are under either IC or LT diffusion model. However, recently, some other diffusion models have also been recently developed, such as Independent Cascade Model with Negative Opinion (IC\mbox{-}N) \cite{chen2011influence}, Opinion cum Interaction Model (OI) \cite{galhotra2016holistic}, Opinion\mbox{-}based Cascading Model (OC) \cite{zhang2013maximizing} etc., which consider negative opinion. SIM Problems and its different variants can also be studied under these newly developed diffusion models.  
\item Most of the studies done on SIM Problem consider that the underlying social network is static including influence probabilities. However, this is not a practical assumption, as most of the social networks are time varying. Recent studies on SIM Problem started considering temporal nature of the social network \cite{tong2017adaptive}, \cite{zhuang2013influence}. As this has just started, there is a lot of scope to work in TSS Problem in time\mbox{-}varying social networks.
\item In real\mbox{-}world social networks, users have specific topics of choice. So, one user will be influenced by other users if both of them have similar choices. Keeping `topic' into consideration spread of influence can be increased, which is known as \textit{topic aware influence maximization}. Recent studies on influence maximization considers this phenomenon \cite{chen2015online} \cite{li2015real}. SIM Problems and its variants can be studied in this settings as well.
\end{itemize} 

\subsubsection{Solution Methodology Specific}
\begin{itemize}
\item Among all the variants of TSS Problem in social networks described in Section \ref{Sec:VTSSP}, it is surprising to see that only SIM problem is well studied. Hence, solution methodologies developed for SIM Problem can be modified accordingly, so that they can be adopted for solving other variants of SIM problem as well.
\item One of the major issues in the solution methodology for SIM problem is the scalability. It is important to observe that the social network used in the Kempe et al.'s \cite{kempe2003maximizing} experiment had 10748 nodes and 53000 edges, whereas the recent study of Nguyen et al.'s  \cite{nguyen2017billion} has used social network of with $41.7 \times 10^{6}$ nodes and $1.5 \times 10^{9}$ edges. From this example, it is clear that the size of the social network data sets is increasing day by day. Hence, developing more scalable algorithms is extremely important to handle large data sets.   
\item From the discussion in Section \ref{Sec:Meta}, it is understood that though there are many evolutionary algorithms, only genetic algorithm, artificial bee colony optimization and discrete particle swarm optimization algorithm have been used till date for solving SIM Problem. Hence, other meta\mbox{-}heuristics, such as \textit{ant colony optimization}, \textit{differential evolution} etc. can also be used for this purpose. 
\item There are many solution methodologies proposed in the literature. However, which one to choose in which situation and for what kind of network structure?  For answering this question, by taking all the proposed methodologies from the literature a strong experimental evaluation is required with benchmark data sets. Recently, Arora et al. \cite{arora2017debunking} has done a benchmarking study with 11 most popular algorithms from the literature, and they have found some contradictions between their own experimental results and reported ones in the literature. More such benchmarking  studies are required to investigate these issues.
\item Most of the algorithms presented in the literature are serial in nature. The issue of scalability in SIM problem can be tackled by developing distributed and parallel algorithms. To the best of the authors' knowledge,   except \textbf{dIRIEr} developed by Zong et al. \cite{zong2014dirier}, there is no distributed algorithm existing in the literature. Recently, a few  parallel algorithms have been developed for SIM Problem \cite{kim2013scalable} \cite{wu2016parallel}. So, this an open area to study the SIM problem and its variants under parallel and distributed settings.
\item Most of the solution methodologies are concerned with the selection of the seeds in one go, before the diffusion starts. In this case, if any one of the selected seeds does not perform up to expectation, then the number of influenced nodes will be lesser than expected. Considering this case, recently the framework of multiphase diffusion has been developed \cite{dhamal2016information}, \cite{han2018efficient}. Different variants of this problem can be studied in this framework.    
\end{itemize}

\section{Concluding Remarks} \label{CR}
 In this survey, first we have discussed the SIM problem and its different variants studied in the literature. Next, we have reported the hardness results of the problem. After that, we have reported major research challenges  concerned  with the SIM Problem and its variants. Subsequently, based on the approach, we have classified the proposed solution methodologies and discussed algorithms of each category. At the end, we have discussed the current research trends and given future directions. From this survey, we can conclude that SIM problem is well studied, though its variants are not and there is a continuous thirst for developing more scalable algorithm for these problems. We hope that presenting three dimensions (variants, hardness results and solution methodologies all together) of the problem will help the researchers and practitioners to have better understanding of the problem and better exposure in this field.
\section*{Acknowledgement}
Authors want to thank Ministry of Human Resource and Development (MHRD), Government of India for sponsoring the project: E-business Center of Excellence under the scheme of Center for Training and Research in Frontier Areas of Science and Technology (FAST), Grant No. F.No.5-5/2014-TS.VII .

\bibliography{paper}

\end{document}